\documentclass[11pt,a4paper]{article}
\pdfoutput=1
\usepackage{jheppub}
\usepackage[utf8]{inputenc}
\usepackage{color,hyperref,dsfont,slashed,multirow}

\graphicspath{{plots/}}

\usepackage{amssymb,amsmath,amsfonts,comment,float,latexsym,graphicx,epstopdf, float}

\usepackage{color}
\usepackage{fancyvrb}
\usepackage{chngcntr}
\usepackage{etoolbox}
\usepackage{booktabs}
\usepackage{bbm,epsfig}
\usepackage{soul}
\usepackage{pdfpages}
\usepackage{array}
\usepackage{lipsum}
\usepackage{amssymb}
\usepackage{graphicx}
\usepackage{hepunits}
\usepackage{slashed}
\usepackage{ccaption}
\usepackage{pdfpages}
\usepackage{scrextend}
 \usepackage{makecell}
\usepackage{cellspace} 
\usepackage[titletoc,toc,title]{appendix}

\hypersetup{bookmarks=true,unicode=true,pdftoolbar=true,pdfmenubar=true,
  pdffitwindow=false,pdfstartview={FitH},pdftitle={},
  pdfauthor={}, pdfsubject={Subject},
  pdfcreator={Creator},pdfproducer={Producer}, pdfkeywords={}{Beyond the Standard Model Physics},
  pdfnewwindow=true,colorlinks=true,
  linkcolor=blue,citecolor=magenta,filecolor=magenta,urlcolor=cyan}
 
\newcommand{\be}{\begin{equation}}
\newcommand{\ee}{\end{equation}}
\def\bsp#1\esp{\begin{split}#1\end{split}}

\newcommand\bpm{\begin{pmatrix}}
\newcommand\epm{\end{pmatrix}}

\def\sectionautorefname~#1\null{Sec.~#1\null}
\def\subsectionautorefname~#1\null{sub--Sec.~#1\null}
\def\figureautorefname~#1\null{Fig.~#1\null}
\def\tableautorefname~#1\null{Table~#1\null}
\def\equationautorefname~#1\null{Eq.~(#1)\null}

\date{\today}

\title{Constraints on the Parameter Space in an Inert Doublet Model with two Active Doublets }

\author[a]{Marco Merchand}
\author[a]{\!, Marc Sher}

\emailAdd{mamerchandmedi@email.wm.edu}
\emailAdd{mtsher@wm.edu}

\affiliation[a]{High Energy Theory Group,  William \& Mary,
Williamsburg, VA 23187, USA}

\abstract{We study a three Higgs doublet model where one doublet is inert and the other two doublets are active. Flavor changing neutral currents are avoided at tree-level by imposing a softly broken $Z'_2$ symmetry and we consider type I and type II Yukawa structures.  The lightest inert scalar is a viable Dark Matter (DM) candidate. A numerical scan of the free parameters is performed taking into account theoretical constraints such as positivity of the scalar potential and unitarity of $2 \rightarrow 2$ scattering amplitudes. The model is further constrained by experimental results such as $B$ physics lower limits on charged Higgs masses, Electroweak Precision Observables, LEP II,  LHC Higgs measurements, Planck measurement of the DM relic abundance and WIMP direct searches by the LUX and XENON1T experiments. The model predictions for mono-jet, mono Z and mono Higgs final states are studied and tested against current LHC data and we find the model to be allowed.    We also discuss the effects of abandoning the ``dark democracy" assumption common in studies of inert models.    Projected sensitivities of direct detection experiments will leave only a tiny window in the DM mass versus coupling plane that is compliant with relic density bounds.    }

\begin{document}
\maketitle


\section{Introduction}
\label{sec:intro}

The nature of dark matter remains one of the biggest mysteries in physics. While evidence for the existence  of DM has been very well established over the last decades, the Standard Model (SM) lacks a good DM candidate. It is therefore necessary to consider theories beyond the SM to address this issue. 

One of the most conspicuous examples of a DM candidate is the so called WIMP whose mass is expected to be of order the electroweak scale $m_{\chi} \approx 100 \text{GeV}$ in order to give the correct annihilation cross section for DM depletion. Models with WIMP candidates exist in abundance in the literature. Perhaps the most famous example is the Minimal Supersymmetric Standard Model (MSSM), however the lack of evidence at the LHC for superpartners has prompted the physics community to look for alternative scenarios.

The inert two Higgs doublet model (IDM) stands as a well motivated non supersymmetric extension of the SM that has within it a viable DM candidate and is still consistent with theoretical and experimental bounds. Like other WIMP models, it predicts monojet, mono-Z, mono-Higgs and vector boson fusion plus missing transverse energy signals at the LHC \cite{Belyaev:2016lok}. 

However the parameter space of the IDM will become more constrained as the LHC program continues improving on its precision measurements of the electroweak sector and as more stringent bounds are placed on the annihilation cross section of DM by direct detection experiments in the upcoming future. Thus it is interesting to consider extensions of the IDM which predict additional phenomena but might evade some of these constraints.

The fact that the IDM doesn't allow CP violation in the scalar sector was one of the main motivations for Grzadkowski, Ogreid and Osland, Ref. \cite{Grzadkowski:2009bt} to extend the IDM by adding an extra active $SU(2)_L$ Higgs doublet.  They called it the IDM2 and  scanned its parameter space imposing theoretical and experimental constraints to determine where DM abundance is acceptable and CP is violated. Although the issue of electroweak baryogenesis was not addressed by the authors they used the difference between the average and the maximal values of the electron electric dipole moment and the basis-independent invariants, introduced by Gunion and Haber in \cite{Gunion:2005ja}, to provide a measure of the amount of CP violation. 

The same model was further studied by some of the same authors in Refs. \cite{Grzadkowski:2010au, Osland:2013sla}. In  \cite{Grzadkowski:2010au} the authors refined the basis invariants used in \cite{Grzadkowski:2009bt} to include the effect from the extra inert doublet and include DM direct detection constraints in their study. In Ref. \cite{Osland:2013sla} the phenomenology of charged scalars at the LHC was studied. 

Another interesting scenario that allows for CP violation is that of a 2HDM plus an inert gauge singlet scalar \cite{Drozd:2014yla}. This model has fewer parameters and the DM is more inert, i.e. it doesn't have gauge interactions. In this model there are two independent portal couplings that allow decoupling between DM annihilation and scattering off nucleons and thus one has to take into account isospin violation i.e. the effective couplings of DM to the proton and the neutron are different and one has to rescale the experimental cross sections. 

The CP conserved version of the IDM2 was studied in ref. \cite{Moretti:2015cwa} by Moretti and Yagyu, together with a model with 2 inert and one active doublet. They referred to these models as I(1+2)HDM  and I(2+1)HDM respectively. They studied the constraints on the parameter space from perturbative unitarity by calculating all possible scalar boson $2 \rightarrow 2$ elastic scatterings. They also included constraints from electroweak precision observables (EWPO) and provided the relevant formulas for the Peskin-Takeuchi S, T and U for both models.

 The results of this paper were used in the I(1+2)HDM by Moretti, Rojas and Yagyu in \cite{Moretti:2015tva} to calculate the one loop induced $H^{\pm} W^{\mp}Z$ vertex and study the parameter space where the branching fraction $H^{\pm} \rightarrow W^{\mp}Z$ can be of order $10 \%$ when the charged scalar is lighter than the top quark.  
  
 The I(2+1)HDM \cite{Keus:2014jha, Keus:2015xya,Cordero:2017owj} has two inert doublets and thus can alleviate the tension with direct detection experiments in the low mass region. In the high mass region, it can bring the model to testable territory by decreasing the mass or increasing the Higgs DM coupling while keeping the required amount of DM relic density. See Refs.   \cite{Keus:2014isa, Ahriche:2015mea, Cordero-Cid:2016krd, Aranda:2019vda} for further studies on this model.

In the past five years or so there has been little investigation of the I(1+2)HDM. An updated revision of the parameter space confronted with the data from run 2 of the LHC would be valuable. It is also important to do a detailed survey of the different mono object signals that arise as predictions of the model and to test them against LHC analyses. 

In this work we study the CP conserving I(1+2)HDM with both type I and type II Yukawa interactions. We take into account theoretical constraints such as positivity of the scalar potential and unitarity constraints on the quartic couplings. B physics bounds on the charged Higgs mass $m_{H \pm}$ as a function of $\tan{\beta}$ are utilized. The most recent LHC Higgs data is enforced as well as the DM relic density results by Planck. Direct detection upper limits on the annihilation cross section as function of DM mass are used to find consistent regions of parameter space. The model was implemented in FeynRules \cite{Alloul:2013bka} and we used micrOMEGAs \cite{Belanger:2013oya,Belanger:2010pz, Belanger:2008sj, Belanger:2006is} to calculate DM observables.

 The outline of the paper is as follows:  the model notation and conventions are introduced in section \ref{model} together with the relevant free parameters. In section \ref{theoretical constraints}
we present the theoretical constraints that will be imposed. Section \ref{experimental constraints} contains the experimental restrictions the model needs to satisfy to be consistent. 
The differences between this model and a simple superposition of the IDM and 2HDM are outlined in section \ref{comparison}. A numerical scan of the parameter space is performed in section \ref{scan}. The predictions of the model for different mono-object final states are examined in section \ref{monoobject}. In Section \ref{NOT-DD}, we discuss the effects of abandoning the ``dark democracy" assumption discussed in the next section; these effects can be substantial in this model.   In section \ref{HeavyHiggs} we discuss constraints from heavy Higgs searches at the LHC.   Section \ref{conclusions} contains our conclusions. We devote two Appendices to include relevant formulas for the 2HDM parameters and for the oblique corrections. 

\section{ Model Description }  \label{model}

\subsection{The inert plus two doublet model} \

The I(1+2)HDM has two active $SU(2)_L$ Higgs doublets that we parametrize as  follows
\begin{equation} 
\Phi_1=\left(
\begin{array}{c}\varphi_1^+\\ (v_1+\rho_1+i\chi_1)/\sqrt{2}
\end{array}\right), \quad
\Phi_2=\left(
\begin{array}{c}
\varphi_2^+\\ (v_2+\rho_2+i\chi_2)/\sqrt{2}
\end{array}
\right),
\end{equation}
while the inert doublet is written as 
\begin{equation}
\eta = \left(
\begin{array}{c}
 \chi^{+} \\ (\chi+i \chi_a)/\sqrt{2} 
\end{array}
\right).
\end{equation}

 The model has a $Z_2\times Z'_2$ symmetry, where the first
factor is the inert-doublet $Z_2$: $\eta \to -\eta$ (all other fields are
neutral) and a softly broken $Z'_2$ is introduced ($\Phi_1 \rightarrow   \Phi_1 , \  \Phi_2 \rightarrow -\Phi_2, $) on the Higgs doublets to avoid tree level FCNC's. 

In this work we follow the notation of Ref.  \cite{Grzadkowski:2009bt} and  we write the potential as 

\begin{equation} \label{Eq:fullpot}
V(\Phi_1,\Phi_2,\eta) 
= V_{12}(\Phi_1,\Phi_2) + V_3(\eta) + V_{123}(\Phi_1,\Phi_2,\eta),
\end{equation}
where $V_{12}$ is the regular 2HDM potential with softly broken $Z_2$, namely 
 \begin{align} 
V_{12}(\Phi_1,\Phi_2) &= -\frac12\left\{m_{11}^2\Phi_1^\dagger\Phi_1 
+ m_{22}^2\Phi_2^\dagger\Phi_2 + \left[m_{12}^2 \Phi_1^\dagger \Phi_2 
+  \text{h.c.} \right] \right\} \nonumber \\
& + \frac{\lambda_1}{2}(\Phi_1^\dagger\Phi_1)^2 
+ \frac{\lambda_2}{2}(\Phi_2^\dagger\Phi_2)^2
+ \lambda_3(\Phi_1^\dagger\Phi_1)(\Phi_2^\dagger\Phi_2) 
+ \lambda_4(\Phi_1^\dagger\Phi_2)(\Phi_2^\dagger\Phi_1) 
\nonumber \\
& + \frac12\left[\lambda_5(\Phi_1^\dagger\Phi_2)^2 + \text{h.c.} \right],
\label{v12} 
\end{align}
while the inert sector potential is simply written as
\begin{equation}
 V_{3}( \eta) = m_\eta^2\eta^\dagger \eta + \frac{\lambda_\eta}{2} 
(\eta^\dagger \eta)^2,
\label{v3} 
 \end{equation}
 and the most general mixing terms between active and inert doublets is given by
 \begin{align}
 V_{123}(\Phi_1,\Phi_2,\eta) 
&=
\lambda_{1133} (\Phi_1^\dagger\Phi_1)(\eta^\dagger \eta)
+\lambda_{2233} (\Phi_2^\dagger\Phi_2)(\eta^\dagger \eta) \nonumber  \\
& +\lambda_{1331}(\Phi_1^\dagger\eta)(\eta^\dagger\Phi_1) 
+\lambda_{2332}(\Phi_2^\dagger\eta)(\eta^\dagger\Phi_2) \nonumber  \\
& 
+\frac{1}{2}\left[\lambda_{1313}(\Phi_1^\dagger\eta)^2 +\text{h.c.} \right]  
+\frac{1}{2}\left[\lambda_{2323}(\Phi_2^\dagger\eta)^2 +\text{h.c.} \right]. \label{Active-Inert}
 \end{align}
 
As we are not interested in investigating electroweak baryogenesis in this scenario we will  assume CP conservation for simplicity and take all the parameters in the scalar sector to be real. We also note that CP non-conservation introduces three mixing angles in the active scalar sector and requires some parameters to be complex.   
A full account of CP violation can be found in Refs.   \cite{Grzadkowski:2009bt,Grzadkowski:2010au}.  

We adopt the "dark democracy" of the quartic couplings
\begin{align} \label{Eq:DarkDemocracy}
\lambda_a\equiv \lambda_{1133}&=\lambda_{2233}, \nonumber \\
\lambda_b\equiv \lambda_{1331}&=\lambda_{2332}, \nonumber \\ 
\lambda_c\equiv \lambda_{1313}&=\lambda_{2323} ,
\end{align} 
this simplification reduces the number of parameters significantly.  As we will see in Section 8, relaxing this assumption, which is common in IDM studies,  can have a substantial effect in this particular model.

The softly broken $Z_2'$ gives rise to four different Yukawa interactions. For a review see Ref. \cite{Branco:2011iw}. In this work we focus on type I model in which all fermions have charge $-1$ under the $Z_2'$ and couple to $\Phi_2$
and the type II model which has down quarks and leptons to be neutral under $Z_2'$ thus coupling to $\Phi_1$ and the up quarks are odd which couples them to $\Phi_2$. The other two types of models called lepton-specific and flipped have identical couplings to quarks as the type I and type II model respectively and are not considered here. 

\subsection{Mass Eigenstates} \label{Parameters}

The diagonalization of the CP odd fields as well as the charged scalars is carried out by the orthogonal transformation (See Appendix \ref{2HDM} for details)
\begin{equation}
\begin{pmatrix}
\chi_1 \\
\chi_2
\end{pmatrix} = \begin{pmatrix}
c_\beta & -s_\beta  \\
s_\beta  &   c_\beta
\end{pmatrix} \begin{pmatrix}
G^0 \\
A
\end{pmatrix}, \quad  \begin{pmatrix}
\varphi_1^{+} \\
\varphi_2^{+}
\end{pmatrix} = \begin{pmatrix}
c_\beta & -s_\beta  \\
s_\beta  &   c_\beta
\end{pmatrix} \begin{pmatrix}
G^{+} \\
H^{+}
\end{pmatrix},
\label{diag1}
\end{equation}
where $c_\beta=\cos \beta$, $s_\beta = \sin \beta$ and $\tan \beta \equiv v_2/v_1$. $G^0$ is the neutral Goldstone boson and $A$ is the physical pseudoscalar while $G^{\pm}$ is the charged Goldstone boson and $H^{+}$ is the charged Higgs.

The physical CP even scalars are obtained by the rotation
\begin{equation}
\begin{pmatrix}
\rho_1 \\
\rho_2 
\end{pmatrix} = \begin{pmatrix}
c_\alpha & -s_\alpha  \\
s_\alpha & c_\alpha
\end{pmatrix}\begin{pmatrix}
H \\
h
\end{pmatrix} , \label{scalars}
\end{equation}
where $h$ and $H$ correspond to the lighter and heavier CP even scalar states respectively.

Notice that there are 8 real parameters in the scalar potential $V_{12}$ and $\tan{\beta} = v_1/v_2$ giving a total of 9 parameters. However the minimization conditions of the potential reduce the number of free parameters down to 7. Here we choose as free parameters the following  
\begin{equation}
S_1 = \left\{ m_h, m_H, m_A, m_{H^{\pm}}, m_{12}^2, \alpha, \beta  \right\}, 
\label{set1}
\end{equation}
which we will call the "active" set as it corresponds to the active Higgs doublets. We write $m_{12}^2$ explicitly in this set as it can have positive or negative values. 

 Since the inert doublet is endowed with a discrete $Z_2$ symmetry its field components do not mix with the Higgs eigenstates and the mass matrices are trivially diagonal in this sector. One can thus solve for the quartic couplings in favor of the squared masses and the quadratic mass term $m_{\eta}^2$ as follows
\begin{equation}
\lambda_a = \frac{2 (m_{\chi \pm}^2 - m_{\eta}^2)}{v^2},
\end{equation}
\begin{equation}
\lambda_b = \frac{m_{\chi_a}^2 - 2 m_{\chi{\pm}}^2 + m_{\chi}^2}{v^2},
\end{equation}
\begin{equation}
\lambda_c = \frac{-m_{\chi_a}^2 + m_{\chi}^2}{v^2}.
\end{equation}
The inert sector $V_{3} + V_{123}$ is thus characterized by 5 parameters 
\begin{equation}
S_2 = \left\{  m_{\chi}, m_{\chi_a}, m_{\chi^{\pm}}, m_{\eta}^2, \lambda_{\eta} \right\}, \label{set2}
\end{equation}
which we call the "inert" set of parameters. The full set of free parameters is thus given by  $S= S_1 + S_2$. 
The SM Higgs boson is fixed to $m_h=125$ $\text{GeV}$  therefore the effective number of free parameters is reduced to 11. Notice that we write $m_{\eta}^2$ explicitly in $S_2$ as it can have positive or negative values.

When the active doublets get a vev the quadratic mass term for the neutral component of the inert doublet is given by 
\begin{equation}
V \supseteq m_{\eta}^2 \eta^{\dagger}\eta +\frac{v^2}{2} \left( \lambda_a \eta^{\dagger}\eta  + \lambda_b |\eta^0|^2  + \lambda_c Re[(\eta^0)^2]  \right),
\end{equation}
therefore in order for $\eta$ not to develop a vev one has to require 
\begin{equation}
m_{\eta}^2 + \frac{v^2}{2} (\lambda_a+\lambda_b+\lambda_c) = m_{\chi}^2 >0, \label{cons1}
\end{equation}
which is automatically satisfied and where we used the expressions for the quartic couplings given above. 


\section{Theoretical Constraints} \label{theoretical constraints}
In this section we provide the theoretical constraints that will be imposed on our numerical scan later in section \eqref{scan}.  These formulas have been derived before, see Refs. \cite{Grzadkowski:2009bt, Moretti:2015cwa}, but we include them here for completeness.

\subsection{Positivity of the potential } \label{Positivity}
The following inequalities involving  the quartic couplings provide the  sufficient conditions for positivity of the scalar potential \cite{Grzadkowski:2009bt}
\begin{align} 
\lambda_1&>0,\quad \lambda_2>0,\quad \lambda_\eta>0,\\
\lambda_x &>-\sqrt{\lambda_1\lambda_2},
\quad \lambda_y>-\sqrt{\lambda_1\lambda_\eta},\quad 
\lambda_y>-\sqrt{\lambda_2\lambda_\eta},\\
\lambda_y &\geq0 \vee \left(\lambda_\eta\lambda_x-\lambda_y^2
>-\sqrt{(\lambda_\eta\lambda_1-\lambda_y^2)
(\lambda_\eta\lambda_2-\lambda_y^2)}\right)
\label{constraints}.
\end{align}
where 
\begin{eqnarray}
\lambda_x&=&\lambda_3+\min\left(0,\lambda_4-|\lambda_5|\right),\\
\lambda_y&=&\lambda_{a}+\min\left(0,\lambda_{b}-|\lambda_{c}|\right).
\end{eqnarray}
In Ref.  \cite{Grzadkowski:2009bt} these conditions were presented as necessary and sufficient. However it has been shown in Ref. \cite{Faro:2019vcd} that these conditions are only sufficient but not necessary. Thus there can be regions of parameter space which violate the conditions above but still have a scalar potential that is bounded from below. In this work we will implement these constraints to guarantee positivity.

\subsection{Unitarity} \label{Unitarity}
The magnitude of the quartic couplings can also be constrained by requiring  unitarity of the S-matrix. The calculation of the s-wave amplitude matrix for all possible $2 \rightarrow 2$ elastic scatterings of scalar bosons for this model was done in Ref. \cite{Moretti:2015cwa}. The requirement for unitarity can be translated into the following conditions
\begin{align}
|x_i|     & < 8\pi,~~(i=1,...9) , \\
|y_j^\pm| & < 8\pi,~~(j=1,...6) , 
\end{align}
where $x_i$ are the eigenvalues of the following matrices 
\begin{equation}
X_1 = 
\begin{pmatrix} 
3\lambda_{\eta} &2\lambda_a + \lambda_b & 2\lambda_a + \lambda_b  \\
 2\lambda_a + \lambda_b & 3\lambda_1 & 2\lambda_3 + \lambda_4\\
  2\lambda_a + \lambda_b & 2\lambda_3 + \lambda_4 & 3\lambda_2 
\end{pmatrix}, ~
X_2  =
\begin{pmatrix} 
\lambda_{\eta} &\lambda_b & \lambda_b  \\
\lambda_b & \lambda_1 &  \lambda_4\\
\lambda_b & \lambda_4 & \lambda_2 
\end{pmatrix},~
X_3 = 
\begin{pmatrix} 
\lambda_{\eta} &\lambda_c & \lambda_c  \\
\lambda_c & \lambda_1 &  \lambda_5\\
\lambda_c & \lambda_5 & \lambda_2 
\end{pmatrix}, \label{3times3}  \\ 
\end{equation}
and 
\begin{align}
y_1^\pm & = \lambda_3 + 2\lambda_4 \pm 3\lambda_5, \label{y1}\\
y_2^\pm & = \lambda_a + 2\lambda_b \pm 3\lambda_c, \\
y_3^\pm & = \lambda_3 \pm \lambda_5, \\
y_4^\pm & = \lambda_a \pm \lambda_c, \\
y_5^\pm & =  \lambda_3 \pm \lambda_4 , \\
y_6^\pm &=  \lambda_a \pm \lambda_b ,
\end{align}
notice that in the most general case there are 18 eigenvalues while in our scenario with the dark democracy assumption the are only 15. The formulas for the most general case are given in Ref. \cite{Moretti:2015cwa}. 

As shown in \cite{Chen:2018uim} several vacuum solutions of the 2HDM may coexist at tree-level and one has to check that the vacuum chosen corresponds to the global minimum of the potential. However in Refs.\cite{Grzadkowski:2010au, Osland:2013sla} it has been mentioned that checking of this restriction is computationally very expensive and that it only eliminates about order $10 \  \%$ of the points in parameter space that satisfy all other restrictions.   Even if one imposes the tree level global minimum conditions, running effects could drive some quartic couplings negative at high energy scales rendering the minimum metastable. A dedicated study of these running effects is beyond the scope of this paper.

\section{Experimental Constraints} \label{experimental constraints}

 \subsection{B physics constraints} \label{Bphys}
 
 Flavor observables, e.g. the b-meson decay $B \rightarrow X_s \gamma$ receive corrections from charged Higgs boson loops and therefore its branching fraction impose constraints on the charged Higgs mass $m_{H\pm}$. The mass of the inert charged Higgs field $\chi^{\pm}$ is innocuous to flavor observables as it doesn't couple to fermions. 
 The most recent fits of the 2HDM to experimental data on flavor physics constraints have been presented in Refs. \cite{Chowdhury:2017aav, Arbey:2017gmh,Haller:2018nnx},  for all four types of Yukawa interactions. 
 
For the type I model we use the most conservative bounds of Ref. \cite{Arbey:2017gmh} while for the type II model we use the $\tan{\beta}$ independent bound $m_{H \pm} > 600$ GeV.

\subsection{Electroweak Precision Observables} \label{EWPO}

The 1-loop corrections to the gauge bosons two-point functions can be encoded by the  Peskin-Takeuchi $S$, $T$ and $U$ parameters, also known as oblique parameters. The difference between the value of this parameters in this model and the SM is written as
\begin{align}
\Delta S[\text{I(1+2)HDM}] = \Delta S_{\text{A}} + \Delta S_{\text{I}}, \notag\\
\Delta T[\text{I(1+2)HDM}] = \Delta T_{\text{A}} + \Delta T_{\text{I}}, \notag\\
\Delta U[\text{I(1+2)HDM}] = \Delta U_{\text{A}} + \Delta U_{\text{I}}. 
\end{align}
where the subscript $A$ stands for the contribution of loop effects due to the active Higgs doublets $\Phi_1$ and $\Phi_2$ while the subscript $I$ stands for the inert doublet $\eta$ contribution, i.e., at 1-loop the effects coming from the active and inert doublets are simply additive.  

The oblique parameters in the general 2HDM were calculated in  \cite{Barbieri:2006dq} and in the IDM in \cite{Bertolini:1985ia}. The formulas for the I(1+2)HDM were presented in Ref.  \cite{Moretti:2015cwa} and we include them in Appendix \ref{EWPOformulas}. 

With $U=0$ fixed, the current measured values of the $S$ and $T$ parameters assuming $m_h=125$ $\text{GeV}$ are given by \cite{Baak:2014ora}
\begin{equation}
\Delta S = 0.06 \pm 0.09, \quad \Delta T= 0.1 \pm 0.07,
\end{equation}
with correlation coefficient $91\%$. For every point of parameter space we calculate the $S$ and $T$ parameters and perform a chi-square test taking into account the correlation coefficient and exclude all points that lie outside the 2$\sigma$ confidence level contour.

\subsection{Constraints from LEP} \label{LEP}
The widths of the gauge $Z$ and $W$ bosons have been measured very precisely at LEP experiments   \cite{Schael:2013ita,TEW:2010aj, Altarelli:1989hv}
. Thus in order to ensure that the decay of the gauge bosons to inert and active scalar sectors are kinematically forbidden we impose the following constraints on the scalar masses
\begin{equation}
2 m_{H^{\pm}}>m_Z, \quad 2 m_{\chi^{\pm}} > m_Z,
\end{equation}
for $Z \rightarrow H^{+}H^{-}$, $\chi^{+}\chi^{-}$,
\begin{equation}
m_A + m_H>m_Z, \quad m_{\chi} + m_{\chi_a}>m_Z,   
\end{equation}
for $Z \rightarrow HA$, $X X_a$ and 
\begin{align}
m_H + m_{H \pm}>m_W,& \quad m_A + m_{H \pm}>m_W, \\
m_{\chi} + m_{\chi \pm}>m_W,& \quad m_{\chi_a} + m_{\chi \pm}>m_W,
\end{align}
for $W^{\pm} \rightarrow H H^{\pm}, A H^{\pm},$ $ \chi \chi^{\pm},\chi_a \chi^{\pm} $.

We also take into account the LEPII MSSM limits applied to the IDM as derived in \cite{Lundstrom:2008ai}. These results exclude the intersection of conditions
\begin{equation}
m_X <80 \ \text{GeV}, \quad  m_{X_a} < 100 \ \text{GeV}, \quad m_{X_a}-m_X >8 \ \text{GeV}.
\end{equation}
Furthermore, LEP collaborations have searched for charged Higgs bosons \cite{Abbiendi:2013hk} and have found no significant excess relative to SM backgrounds. In this work we take the conservative lower bound  
\begin{equation}
m_{H^{\pm}}, m_{X^{\pm}} > 70 \ \text{GeV},
\end{equation}
as was found in Ref. \cite{Pierce:2007ut}, see also Ref. \cite{Arbey:2017gmh}.

\subsection{LHC Higgs data} \label{LHC}

As discussed in section \eqref{model}, the I(1+2)HDM is determined by 12 free parameters. However we fix the value of $m_h=125$ $\text{GeV}$ such that the field $h$ corresponds to the SM Higgs boson. 
We also notice that the Higgs boson couplings to vector bosons normalized to the SM value are $g_{hVV} = \sin{(\beta-\alpha)}$ and measurements of this coupling at the LHC are very consistent with the SM value of $g_{hVV}^{\text{SM}} =1$ at the level of $10 \%$ and is expected to improve at the HL-LHC \cite{Dawson:2013bba} to about $2\%$ accuracy.    

In this work we focus on the case where $\chi$ is the lightest particle in the inert sector, i.e. $M_{\chi}<M_{\chi_a}$ and $M_{\chi}<M_{\chi^{\pm}}$. For light enough dark matter, $M_{\chi}<m_h/2$ the invisible decay channel of the Higgs boson is kinematically open and is given by
\begin{equation}
\Gamma (h \rightarrow \chi \chi) = \frac{(\lambda_a+ \lambda_b+ \lambda_c)^2}{32 \pi} \frac{v^2}{m_h}\sin^2{(\beta-\alpha)}\sqrt{1-\frac{4 M_{\chi}^2}{m_h^2}}. \label{h_to_inv}
\end{equation}

Constraints on the Higgs boson branching ratio to invisible final states have been  reported by the ATLAS collaboration to be $BR(h \rightarrow \text{invisible}) < 28 \%$ at the $95 \%$ C.L. The most recent constraint has been reported by CMS group and is given by \cite{Sirunyan:2018owy}
\begin{equation}
BR(h \rightarrow \text{invisible}) < 19 \%,
\end{equation}
at the $95 \%$ C.L. We will use this bound in our numerical scan. 

It is well known that in the 2HDM the effective coupling of the Higgs bosons to pairs of photons receive contributions from charged scalar loops. In the I(1+2)HDM the Higgs couplings to photons receive extra contributions from loops of inert charged Higgs $\chi^{\pm}$. We have implemented the effective $gg$ and $\gamma \gamma$ couplings of the Higgs bosons in micrOMEGAs.

To take into account the experimental constraints from all different Higgs signal strengths measured at the LHC and the Tevatron we have used the Lilith library \cite{Bernon:2015hsa}. In addition, exclusion limits from heavy Higgs searches are taken into account by use of the HiggsBounds \cite{Bechtle:2011sb} code.

\subsection{Relic Density}

The latest results from the Planck collaboration \cite{Aghanim:2018eyx} give the following value for the DM relic density
\begin{equation}
\Omega_{\text{DM}} h^2 = 0.120 \pm 0.001.
\end{equation}

 In the numerical scan we allow the model to predict the dark matter under-abundance as other field components could contribute to the relic density and impose the upper bound as an experimental constraint. The relic density $\Omega_{\text{DM}}^{\text{Planck}} h^2$ was evaluated with the micrOMEGAs package \cite{Belanger:2013oya}. The annihilation into three body final state with a virtual $Z$ and $W$ bosons were included in the calculation of the relic density.
 
We assume $10 \%$ theoretical uncertainty as the calculation of the relic density is performed at tree-level. Thus we inflated the experimental error to $10 \%$ of the central value and use two standard deviations to set limits 
\begin{equation}
\Omega^{\text{limit}}_{\text{DM}} h^2 =0.120 \pm 2 \times 0.012. \label{reliclimit}
\end{equation}

\subsection{Direct detection experiments} \label{Relic}

Experiments such as LUX \cite{Akerib:2016vxi}, PANDAX-II \cite{Cui:2017nnn} and the XENON1T \cite{Aprile:2018dbl} place constraints on the spin-independent cross section of Weakly interacting Massive Particles (WIMP) off nucleons as a function of the WIMP mass. 

We have used the micrOMEGAs package \cite{Belanger:2013oya} to evaluate the spin-independent cross section for DM scattering off a proton.
To constraint the model parameters we rescale the cross section by a factor $\Omega_{\text{DM}}/\Omega_{\text{DM}}^{\text{Planck}}$ that takes into account that $\chi$ represents only a part of the total DM.  

To calculate the spin independent (SI) amplitude micrOMEGAs calculates the effective coupling of the DM candidate with quarks and automatically takes into account loop contributions from box diagrams following the model independent calculation of Ref. \cite{Hisano:2015bma}. The DM-quark amplitudes are related to the DM-nucleon amplitudes by form factors that are stored as global parameters. We have checked that the effect of isospin violation is negligible in this scenario in contrast to the 2HDM plus a gauge singlet scalar  \cite{Drozd:2014yla} where this effect was found to be negligible in the type-I model allowing the direct use of the experimental upper limits, while it was more significant for the type-II.

We note that there are two loop triangle diagrams with either $h$ or $H$ being exchanged.  In the alignment limit $\sin{(\beta-\alpha)} =1$ only $h$ exchange is supported and the coupling strength scales as 
\begin{equation}
\lambda_{abc} \equiv \lambda_a+ \lambda_b+ \lambda_c = 2\frac{m_{\chi}^2-m_{\eta}^2}{v^2}, \label{DMcoupling}
\end{equation}
therefore the DM-nucleon interaction is determined by the difference between the DM mass and the quadratic mass term of the inert doublet potential. This is the same scaling behavior of the invisible decay width of the Higgs boson, see equation \eqref{h_to_inv}. Contrary to a naive expectation, the potential mass term $m_{\eta}$ is phenomenologically relevant as for a given DM mass value it moderates several DM observables. 

In the IDM the bounds from indirect detection experiments, e.g. AMS-02 or Fermi-LAT are much weaker than the LHC and direct detection constraints above, see Refs.   \cite{ Eiteneuer:2017hoh,Arhrib:2013ela}. Hence we do not consider constraints coming from direct detection experiments in this work.

\section{Comparison with superposition of IDM + 2HDM }  \label{comparison}
 Although some constraints e.g.  the change in the Peskin-Takeuchi parameters at leading order, are a simple sum of the active and inert sector contributions, it must be  emphasized that in general the I(1+2)HDM is not just a simple superposition of the regular IDM and 2HDM. 
 
 The addition of an extra active doublet to the regular IDM gives rise to notable differences from just the regular IDM or 2HDM. These differences are encoded in the quartic couplings which parametrize the interaction between the two sectors and are given in eqs. \eqref{Eq:DarkDemocracy}. The most important ones in this paper are the following:
 \begin{itemize}
 
\item{Positivity and Unitarity.} At the theoretical level, the quartic couplings lead to non-trivial positivity as well as unitarity conditions which are not just the sum of conditions given in the IDM and 2HDM, see sections \ref{Positivity} and \ref{Unitarity}. 

\item{Effective coupling to photons.} The Higgs coupling to a pair of photons receives loop contributions from both active and inert charged Higgs states. Depending on the sign and size of the quartic and trilinear Higgs couplings there can be cancellations or enhancements between the two diagrams, an effect that is not present in either the IDM or 2HDM.

\item{Invisible decays and DM annihilation.}  The invisible decay channel of Higgs states as well as the decay mode into charged inert states $h,H \rightarrow \chi^+ \chi^-$  arise naturally in this scenario while they are absent in just the 2HDM.  These decay modes are further controlled by the mixing angles of the active sector so that $g_{h\chi \chi} \propto \lambda_{abc} \sin{(\beta - \alpha)}$ and $g_{H\chi \chi} \propto \lambda_{abc} \cos{(\beta - \alpha)}$ hence in the alignment limit only the SM Higgs is allowed to decay invisibly. This also affects the amplitude of DM annihilation via the Higgs mediated diagrams which is important mostly in the low mass region.  The are also quartic couplings which are independent of the mixing angles, namely 
\begin{equation}
g_{\chi \chi hh} = g_{\chi \chi HH} = \lambda_{abc}, \quad  g_{\chi \chi AA} = \lambda_{a} +\lambda_{b}-\lambda_{c},
\end{equation}
 which are mainly relevant in the high mass region. 

\item{Mono object production.}  The cross sections for mono-jet, mono Z and mono Higgs final states pick up contributions from the active doublet states which are not present in the IDM alone. Although for mono-jet and mono Z  these effects turn out to be negligible for the total cross section, they can be substantial for mono Higgs final states which are also controlled by trilinear Higgs couplings coming from the 2HDM exclusively.

\end{itemize} 

\section{Numerical Scan of Parameter Space} \label{scan}
We perform a random scan of the free parameters of the model, according to the following ranges
\begin{equation}
2 \leq \tan{\beta} \leq 10,
\end{equation}
\begin{equation}
0  \leq \beta - \alpha \leq   \pi,
\end{equation}
\begin{equation}
10 \  \text{GeV} \leq m \leq 1000 \ \text{GeV}, \quad \text{with} \  m=m_{H}, m_{A},  m_{\chi}, m_{\chi_a},
\end{equation}
\begin{equation}
m_Z/2 \  \text{GeV} \leq m \leq 1000 \ \text{GeV}, \quad \text{with} \  m= m_{H \pm}, m_{\chi \pm},
\end{equation}
\begin{equation}
-1 \ \text{TeV}^2 \leq m_{12}^2 \leq 1 \ \text{TeV}^2,
\end{equation}
\begin{equation}
-1  \ \text{TeV}^2 \leq m_{\eta}^2 \leq  1 \ \text{TeV}^2,
\end{equation}
\begin{equation}
0  \leq \lambda_{\eta} \leq  8 \pi.
\end{equation}

The scan is done in a succession of cuts as follows: 
\begin{enumerate}
\item Cut 1. We first random scan the parameters imposing the positivity,  unitarity and B physics constraints of sections \eqref{Positivity}, \eqref{Unitarity} and \eqref{Bphys} respectively. The conditions $m_{\chi} < m_{\chi_a}$ and $m_{\chi} < m_{\chi \pm}$ are implemented to make sure $\chi$ is the lightest inert scalar. We also avoid the degenerate regime where the CP even active scalar has mass in the range $m_H \ \in \ [123, 128]$ GeV.

\item Cut 2. In this cut we evaluate a chi-square function giving the $2 \sigma$ confidence region where EWPO are satisfied and impose LEP constraints given in sections \eqref{EWPO} and \eqref{LEP} respectively.

\item Cut 3. We apply the constraints coming from the Higgs boson signal strengths and heavy Higgs searches given in section \eqref{LHC}. We discard all points that give DM overabundance see equation \eqref{reliclimit}.

 \item Cut 4. Finally we throw away all points in the parameter space that produce spin-independent cross section above the quoted limits by LUX and XENON1T experiments. The bounds from PANDAX-II experiment are less severe and thus are not imposed. Data from XENON1T was taken from \cite{PhenoData}.

 \end{enumerate}

\begin{figure}[H]
\centering
\includegraphics[scale = 0.49] {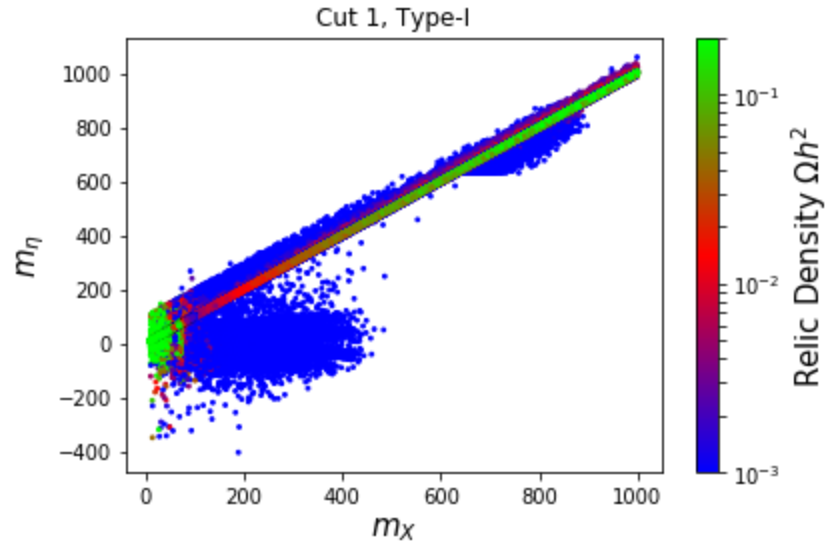}
\hspace{.0009mm}
\includegraphics[scale = 0.49] {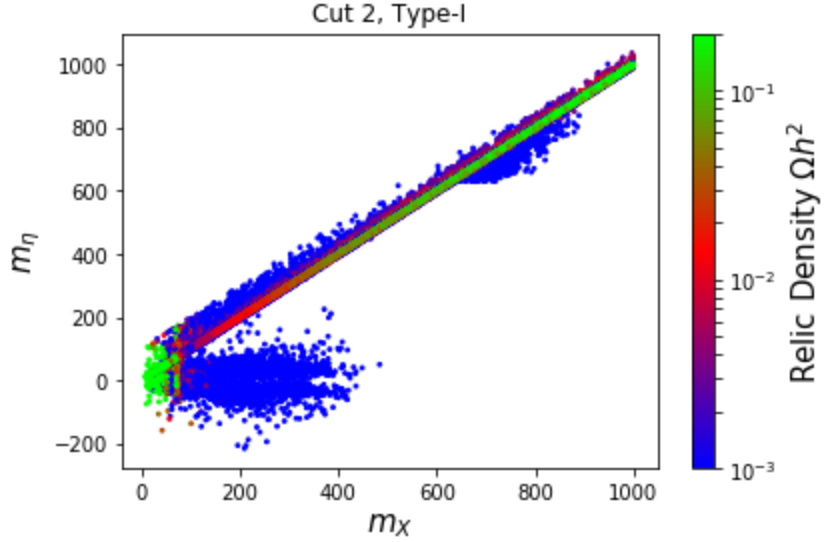}
\hspace{.1mm}
\includegraphics[scale = 0.49] {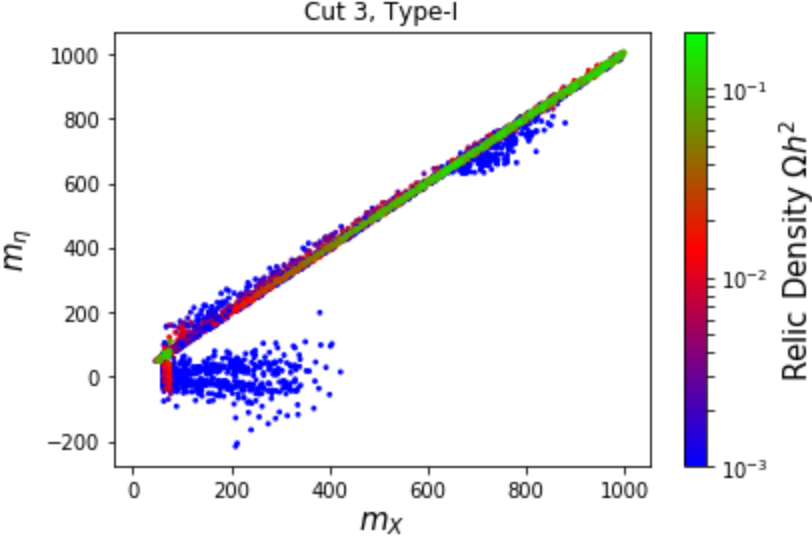}
\hspace{.1mm}
\includegraphics[scale = 0.49] {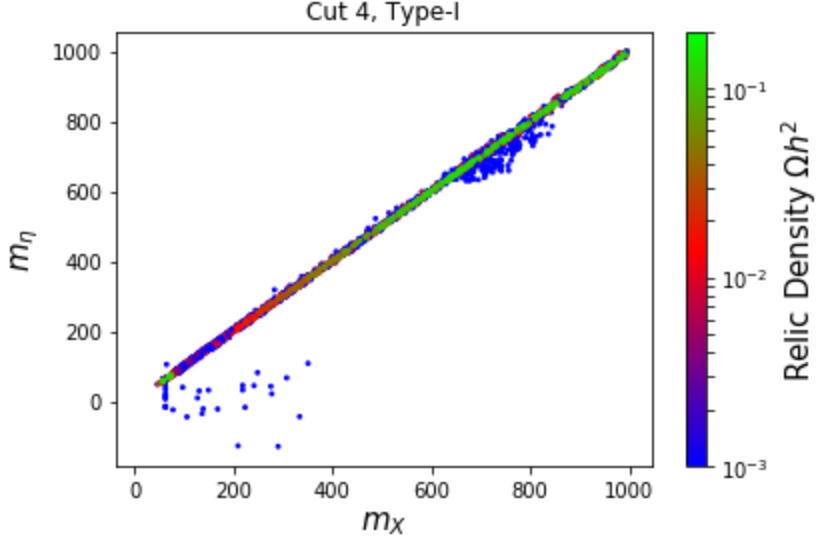}
\caption{ Allowed points of parameter space that survived the succession of cuts. In Cut 1 (upper left) we applied unitarity, positivity and B physics constraints, in Cut 2 (upper right) shows points that survive after imposing EWPO and LEP constraints, in Cut 3 (lower left) the LHC constraints on the Higgs boson signal strengths and heavy Higgs searches limits were imposed as well as the upper bound on DM relic abundance are applied and finally in  Cut 4 (lower right) the LUX and XENON1T experimental results on the spin independent cross section were implemented. In the vertical axis we took the square root of the absolute value of inert sector mass parameter $m_{\eta}$ so that negative allowed values correspond to $-m_{\eta}^2$ in the scalar potential.  }\label{ALLCUTS}
\end{figure}

   There is almost no correlation between the active and inert parameter sets, $S_1$ and $S_2$, that we introduced in section \eqref{Parameters} thus we choose to present the results of the parameter scan in figure \ref{ALLCUTS} in terms of the most relevant parameters that affect the relic density, namely the DM mass $m_{\chi}$ and the quadratic mass term of the inert sector $m_{\eta}$.  From the plots it is evident that as the various parameter scan cuts are applied the $m_{\chi}$ and $m_{\eta}$ parameters are forced to become more degenerate. The reason can be seen from Eq. \eqref{DMcoupling}, where one can see that the degeneracy is a result of $\lambda_{abc}$ not being too large.

During the scan we only imposed the Planck upper limit \eqref{reliclimit} on the relic density as there can be other components or dark sectors that contribute to DM production.  The most salient result from these figures is the identification of two regions where the experimental lower bound on the relic density is also  satisfied. There is a low mass region with DM mass in the range $[ 57, 73 ]$ GeV and a high mass region with $m_{\chi }$ in the range $[ 500, 1000]$ GeV. These points are highlighted in green on the figure.  We will refer to these low and high DM mass regions compliant with Planck limits as the low-DM region (LDM) and high-DM region (HDM). We want to note here that the identification of the LDM and HDM regions agrees with the parameter space study of Ref.  \cite{Grzadkowski:2010au}, see figure 2 in that reference. The plots presented in figure \ref{ALLCUTS} were generated for the type-I model. For the type II model no qualitative difference arises therefore the plots are not presented.  

\begin{figure}[H]
\centering
\includegraphics[scale = 0.49] {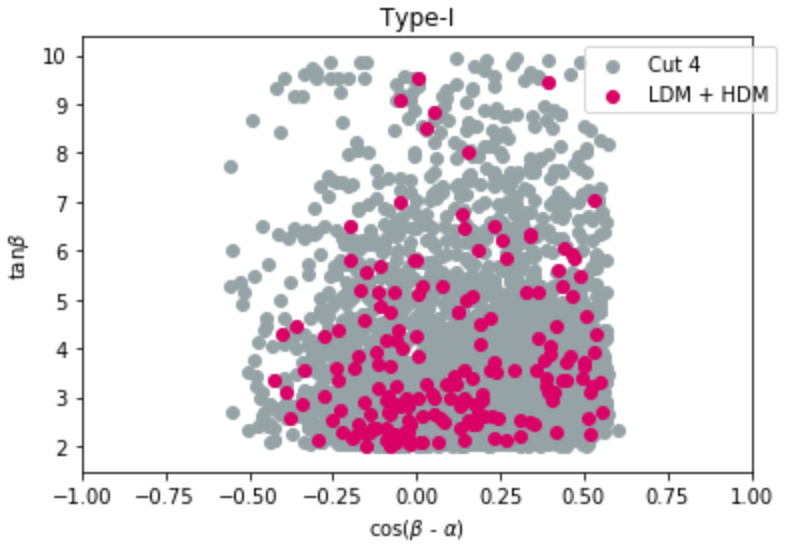}
\hspace{.0029mm}
\includegraphics[scale = 0.49] {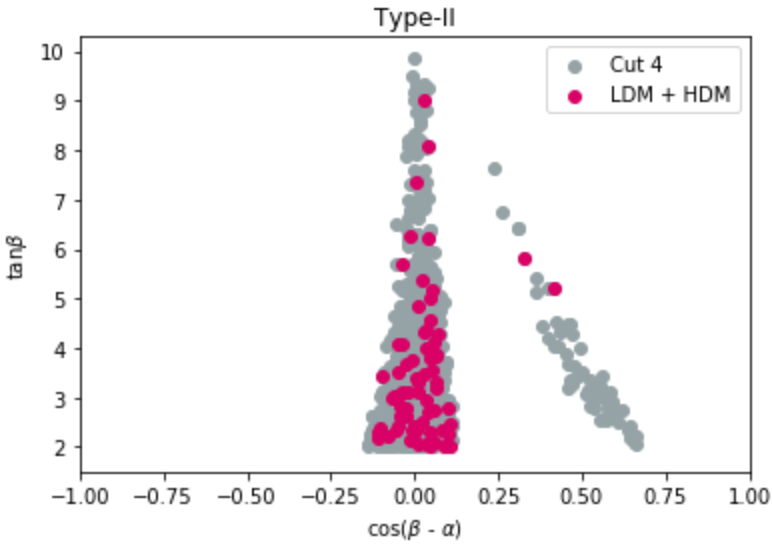}
\hspace{.0009mm}
\includegraphics[scale = 0.49] {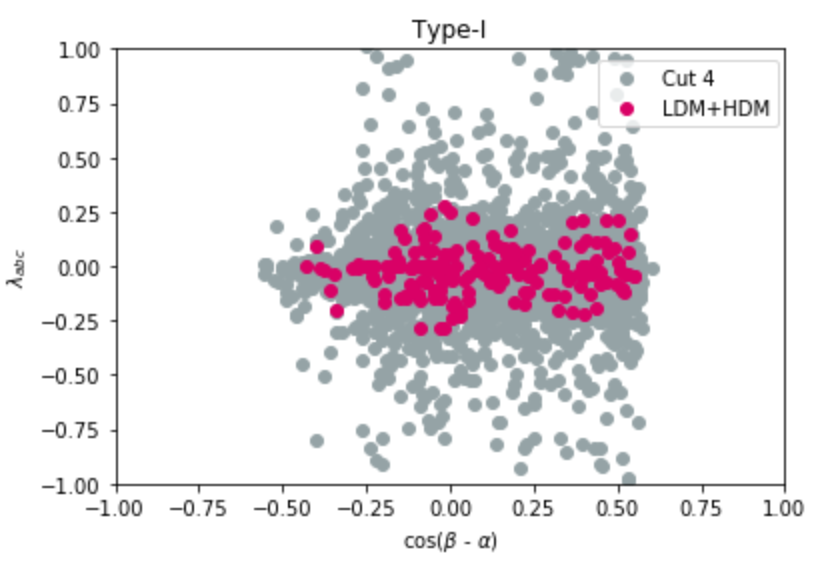}
\hspace{.0009mm}
\includegraphics[scale = 0.49] {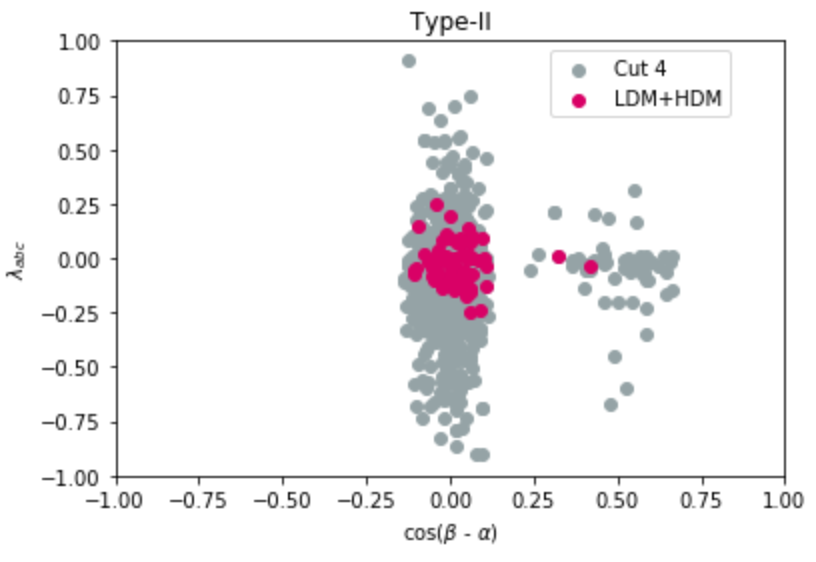}
\caption{ Parameter space points in the $\cos{(\beta-\alpha)}$-$\tan{\beta}$ (upper) and  $\cos{(\beta-\alpha)}$-$\lambda_{abc}$ (lower) planes. All points survived the cut 4 of the parameter scan while the pink dots are also compliant with Planck lower limit. }\label{alpha_beta}
\end{figure}

The allowed values for the mixing angles in the active sector are mainly constrained by the measured Higgs signal strengths. We present scatter plots for the $\cos{(\beta - \alpha)}$-$\tan{\beta}$ parameter space consistent with the parameter scan in figure \ref{alpha_beta}. Superimposed and colored green are the points that also comply with Planck lower limits on the relic density. One can notice that DM constraints have a moderate impact on the parameter space of the mixing angles. The upper panel plots agree with the overall shape from the latest 2HDM fit presented in Ref. \cite{Haller:2018nnx}. Also shown is the $\cos{(\beta-\alpha)}$-$\lambda_{abc}$ plane which shows that relic density constraints are not affected by the allowed range of $\cos{(\beta - \alpha)}$ but only by the absolute value of the portal interaction $\lambda_{abc}$.

\begin{figure}[H]
\centering
\includegraphics[scale = 0.5] {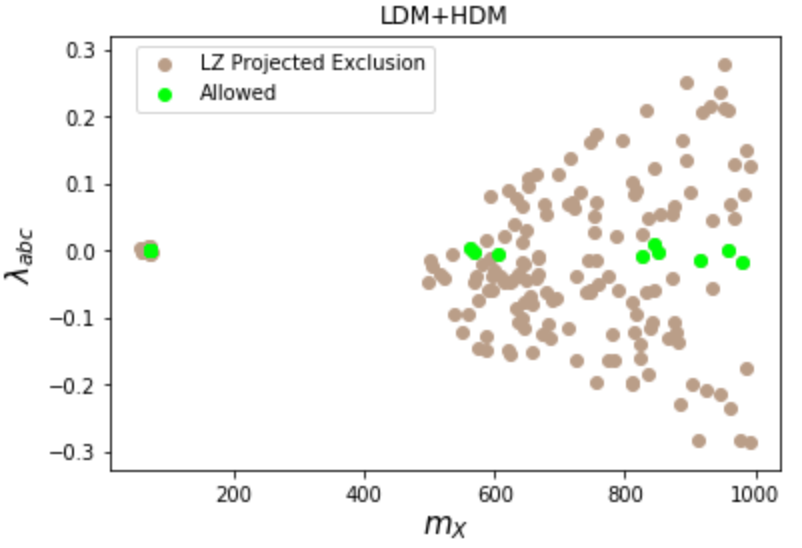}
\includegraphics[scale = 0.5] {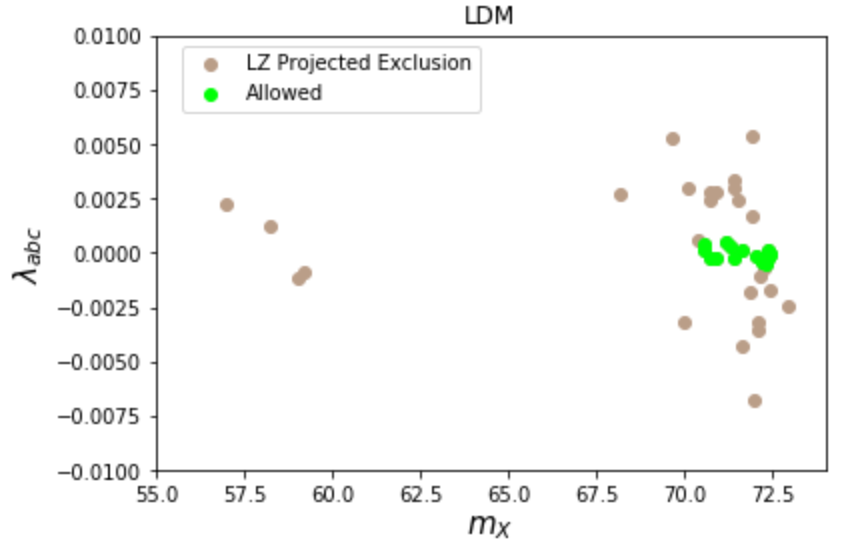}
\caption{  LDM+HDM (left) and LDM (right) regions where points colored brown would be excluded by LZ projected upper limits while points colored green would still be allowed.} \label{LDM_HDM}
\end{figure}

When the projected upper bounds from the LZ collaboration \cite{Akerib:2015cja} are applied to the "LDM+HDM" data set  we find that the allowed parameter space can be dramatically diminished sending the model to a fine tuned region where $|\lambda_{abc}| \leq 5 \times 10^{-4}$ (corresponding to $|m_{\eta} - m_{\chi}| \leq 0.1$ GeV  ) in LDM and   $|\lambda_{abc}| \leq 0.02$ (corresponding to $|m_{\eta} - m_{\chi}| \leq 0.25$ GeV) in HDM. This can be appreciated in figure \ref{LDM_HDM} which shows the $m_{\chi}$-$\lambda_{abc}$ parameter space points for LDM+HDM and LDM. 

\begin{figure}[H]
\centering
\includegraphics[scale = 0.49] {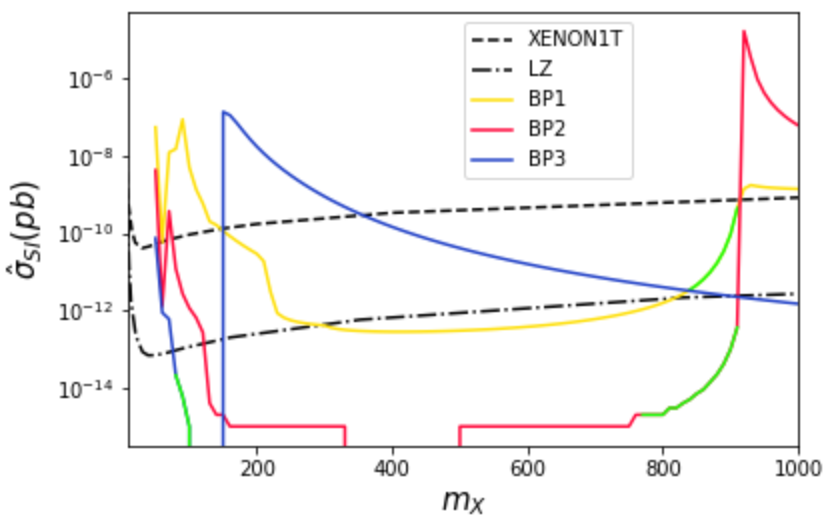}
\caption{  DM spin independent scattering cross section with nucleons as function of $m_{\chi}$ for benchmark points of table \ref{benchmarks1}. The upper limits from XENON1T (dashed) and projected from LZ (dashed-dotted) experiments are shown for comparison. For each curve, the green segments are allowed by all theoretical and experimental constraints. }\label{SIcs}
\end{figure}

One can also observe that for the LDM region there would be a tiny window that survives with $70 \leq m_{\chi} \leq 72.5$ GeV. 
The effects of the XENON1T and projected LZ bounds can be further appreciated in figure \ref{SIcs} where we plot the spin independent scattering cross section with nucleons as a function of the DM mass keeping the portal interaction fixed, i.e. for each value of $m_{\chi}$ we fixed $m_{\eta}$ such that $\lambda_{abc}$ remains fixed. We chose three benchmark points: BP1 with the maximum absolute value for $\lambda_{abc}$ and BP2 and BP3 which where drawn from the LDM+HDM region that would not be ruled out by the projected LZ bounds, corresponding to green colored points in figure \ref{LDM_HDM}.  The rest of the free parameters remained fixed and the benchmark points are displayed in table \ref{benchmarks1}. Since for each benchmark value chosen we are varying the DM mass, we indicate with color green on each curve the segment that remains consistent with all the constraints. 

\begin{table}[h]
\begin{center}
\begin{tabular}{ | m{.9 cm} |m{.9cm}| m{1.2 cm}| m{.9 cm} | m{0.6 cm} | m{0.7 cm} |  m{0.7 cm} | m{0.6 cm} |m{1.005 cm} | m{0.6 cm} | m{1.7cm} | } 
\hline
 & $\tan{\beta}$ &  $c_{\beta - \alpha}$&$m_{12}$  & $m_H$ & $m_A$ & $m_{H \pm}$  & $m_{ \chi a}$ & $m_{ \chi \pm}$ & $\lambda_{\eta}$ & $\lambda_{abc}$ \\ 
\hline
BP1 & $2.06 $ & $ -0.02$& $49$ & $261$ & $307$ & $ 222$ & $919$  & $920$ &$6$ &$-0.28$\\ 
\hline
BP2 & $2.56$ & $-0.38$& $91$ & $295$ & $129$ & $ 281$ &$924$ & $920$ &$0.32$ &$-0.015$ \\ 
\hline
BP3 & $2.21$ & $-0.08$& $-225$ & $82$ & $100$ & $ 203$ &$363$ & $142$ &$3.25$ &$-4\times 10^{-4}$ \\ 
\hline
\end{tabular}
\end{center} \caption{Benchmark points as drawn originally from the numerical scan which correspond to the curves of figure \ref{SIcs} .We took the square root of the absolute value of active sector mass parameter $m_{12}$ so that a negative value corresponds to $-m_{12}^2$ in the scalar potential. All mass parameters are in GeV units. } \label{benchmarks1}
\end{table}

\section{Model Predictions} \label{monoobject}

 We devote this section to studying mono object signals plus missing transverse energy and the prospects for the LHC to probe them. Of particular interest are processes with a jet, a $Z$ boson or a Higgs boson in the final state. For the rest of this section, unless otherwise noted, whenever we refer to the allowed parameters in our scan we imply all points that fall into the whole LDM $+$ HDM region where relic density is fully accounted by the inert sector of this model. The study of signals at lepton colliders in this model are left for future work. In the IDM this has been investigated in Refs. \cite{Kalinowski:2018ylg,Kalinowski:2018kdn,arXiv:1911.06254}.

In the calculation of the production rates we implemented the following configurations and cuts:
\begin{itemize} 
\item[1] For the proton initial states we used the PDF NNPDF23$\_$lo$\_$as$\_$0130$\_$qed \cite{Ball:2013hta,Ball:2012cx}
that is implemented within CalcHEP. 

\item[2] The QCD renormalization and factorization scales were set equal to the missing transverse momentum of the final states for all processes. 

\item[3] A minimum cut of $100$ GeV is placed on the missing transverse momentum for all processes.

\item[4] For all processes, the proton and jets have been defined as composite states of a gluon and all quarks except the top.

\end{itemize} 

The calculation of the cross sections is performed by CalcHEP at leading order in perturbation theory.  In this work, we are not including higher order corrections.    In a detailed study of the inert doublet model \cite{Dercks:2018wch}, these corrections were included, including a discussion of the choice of renormalization and factorization scales and the K-factors for the production cross sections.   This model, however, has many more parameters than the simple inert doublet model, some of which (such as heavy scalar masses) will not be measured in the near future, and thus it would seem premature to include these effects.  Including them would not have a substantial qualitative effect on our results. 

 What are the experimental bounds and prospects? A very detailed analysis is beyond the scope of this paper, and would also depend on several extra parameters.   However one can get a rough idea from the IDM analyses in recent works of Refs. \cite{Dercks:2018wch} and \cite{Belyaev:2018ext}.    The latter paper gives approximate bounds from Run 2 (their Figure 13) and show that the upper bounds are one to two orders of magnitude above the expected value in the IDM.    Our results are the same order of magnitude for light $m_\chi$ (which is the expected region if $\chi$ is the dark matter).   After $3000\ {\rm fb}^{-1}$ at the HL-LHC, the expected reach becomes comparable to the IDM results for low masses, and the same is true in our case.   The same is true for the mono Z signature in the next section.    Of course, should an additional active doublet be detected, a much more detailed analysis would be in order.

\subsection{Mono jet}

In addition to Ref. \cite{Dercks:2018wch}, mono jet signals have been studied in the context of the inert two Higgs doublet model, see e.g. Refs.   \cite{Belyaev:2016lok,Belyaev:2018ext,Arhrib:2013ela}, see also Ref. \cite{Craig:2014lda} for related studies in other Higgs portal scenarios. 
 In a similar way, the I(1+2)HDM predicts two different kinds of jet plus missing transverse energy final states. The two possibilities are  $p p \rightarrow j \chi \chi$ and  $p p \rightarrow j \chi \chi_a$ respectively. 
 
 The process $p p \rightarrow j \chi \chi$ is determined by the couplings of the Higgs eigenstates to fermions which control the production cross section by gluon fusion. The most important contribution comes from top quark loops and the couplings to quarks scale as $g_{htt}$, $g_{Htt} \propto 1/\tan{\beta}$ for both type-I and -II models.

  The other parameter affecting the mono-jet production is given by the interaction of the Higgs eigenstates with the DM particle which goes as $\lambda_{abc} \sin{(\beta - \alpha)}$ for $h$ and as $\lambda_{abc} \cos{(\beta - \alpha)}$ for $H$ where $\lambda_{abc}\equiv \lambda_a +  \lambda_b + \lambda_c$ was defined in eq. \eqref{DMcoupling}. Therefore in the alignment limit only the SM Higgs will mediate this process. 
 
Thus we expect that the biggest effects will come from the maximum absolute value allowed for the Higgs DM coupling and for small values of $\tan{\beta}$.  We have chosen three benchmark points (see table \ref{monojet_XX_benchmarks}) with maximum absolute value of $\lambda_{abc} \approx 0.2$ for three different values of $\cos{(\beta-\alpha)}$. Cross sections for the process $p p \rightarrow j \chi \chi $ are presented in figure \ref{monojetXXplot} below.  For BP4 we chose the minimum value of $\cos{(\beta-\alpha)}$ for comparison.  Since in type-II model the deviations from alignment limit are more stringent, we present cross sections for the type-I only.

  \begin{table}[h]
\begin{center}
\begin{tabular}{ | m{.9 cm} |m{1.cm}| m{1.4 cm}| m{.9 cm} | m{0.8 cm} | m{0.8 cm} |  m{0.8 cm} | m{0.6 cm} |m{0.905 cm} | m{0.6 cm} | m{1.cm} | } 
\hline
 & $\tan{\beta}$ &  $c_{\beta - \alpha}$&$m_{12}$  & $m_H$ & $m_A$ & $m_{H \pm}$  & $m_{ \chi a}$ & $m_{ \chi \pm}$ & $\lambda_{\eta}$ & $\lambda_{abc}$ \\ 
\hline
BP4 & $4.26 $ & $ 6 \times 10^{-4}$& $-80$ & $108$ & $407$ & $ 88$ & $894$  & $895$ &$3.48$ &$0.25$\\ 
\hline
BP5 & $3.65$ & $0.5$& $250$ & $358$ & $315$ & $ 380$ &$834$ & $836$ &$7.8$ &$0.21$ \\ 
\hline
BP6 & $3.55$ & $-0.34$& $86$ & $256$ & $67$ & $ 238$ &$818$ & $815$ &$5.78$ &$ -0.20$ \\ 
\hline
\end{tabular}
\end{center} \caption{Benchmark points as drawn originally from the numerical scan which correspond to the curves of figure \ref{monojetXXplot} .We took the square root of the absolute value of active sector mass parameter $m_{12}$ so that a negative value corresponds to $-m_{12}^2$ in the scalar potential. All mass parameters are in GeV units. } \label{monojet_XX_benchmarks}
\end{table}

\begin{figure}[H]
\centering
\includegraphics[scale = 0.5] {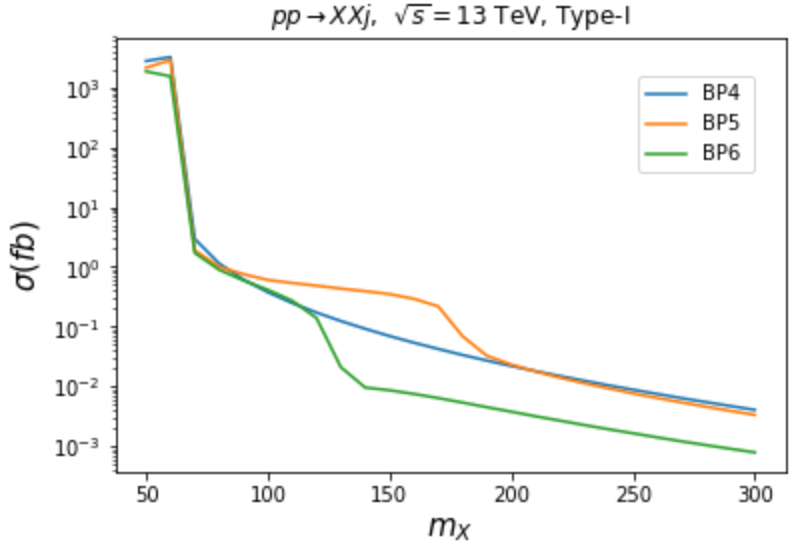}
\caption{ Cross sections for mono jet production $j \chi \chi$  as a function of DM mass for the benchmark points presented in table \ref{monojet_XX_benchmarks}. The mass squared parameter $m_{\eta}^2$ was also varied so that the portal coupling $\lambda_{abc}$ remained fixed as displayed on the table.} \label{monojetXXplot}
\end{figure}

\begin{figure}[H]
\centering
\includegraphics[scale = 0.5] {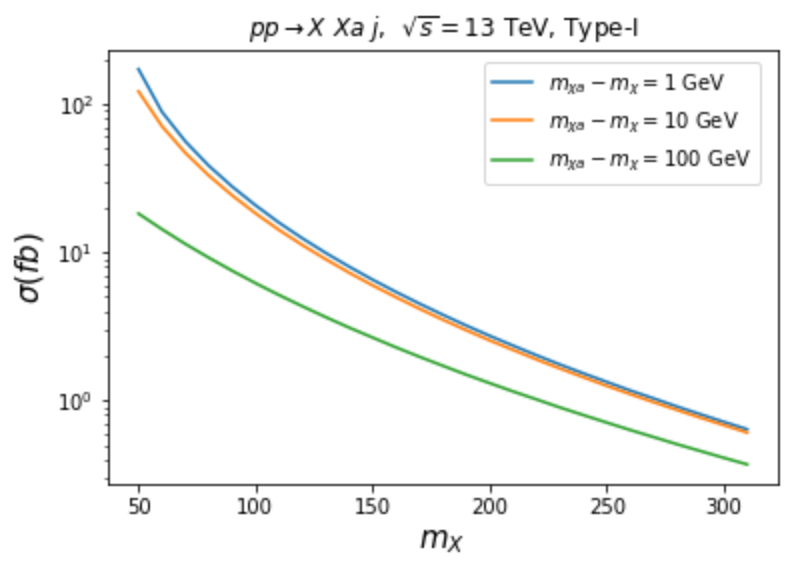}
\caption{ Cross sections for the process $ p p \rightarrow j \chi \chi_a$ for different mass separations.} \label{monojetXXa}
\end{figure}

The other possibility for monojet final state is given by the process $ p p \rightarrow j \chi \chi_a$. This is even simpler to describe as the only relevant parameters are the DM masses $m_{\chi}$ and $m_{\chi a}$. A $Z$ boson is mediated in this process and the $Z \chi \chi_a$ vertex is fixed by electroweak parameters. The effect is more significant for small mass separation. This effect can be appreciated in figure \ref{monojetXXa} where we plot the cross section as a function of $m_{\chi}$ for different mass separations.

\subsection{Mono Z}
\begin{figure}[H]
\centering
\includegraphics[scale = 0.34] {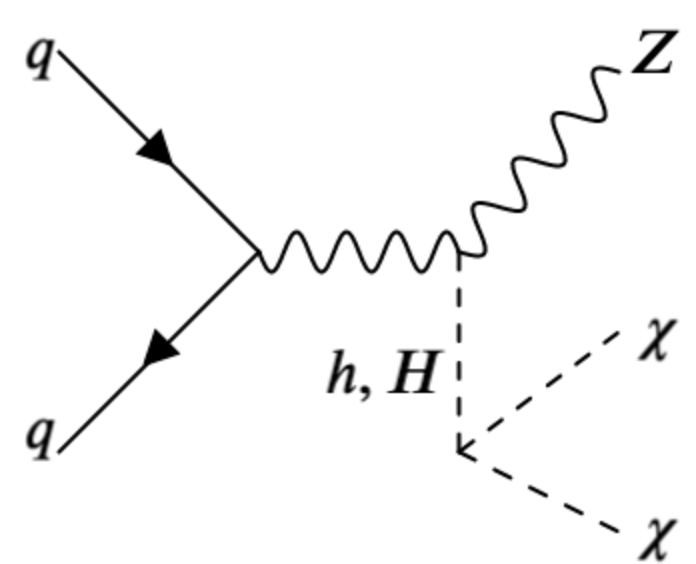}
\hspace{.034mm}
\includegraphics[scale = 0.34] {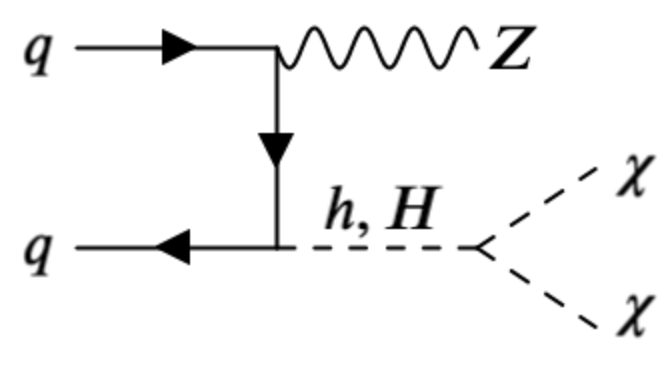}
\hspace{.034mm}
\includegraphics[scale = 0.34] {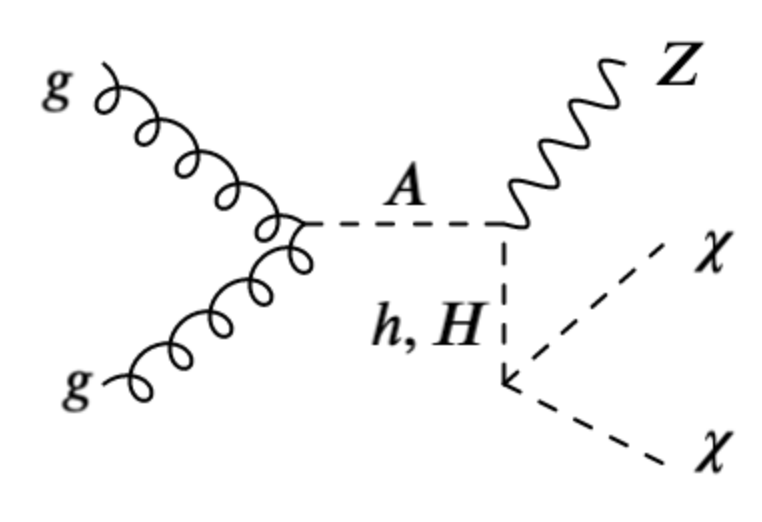}
\hspace{.034mm}
\includegraphics[scale = 0.34] {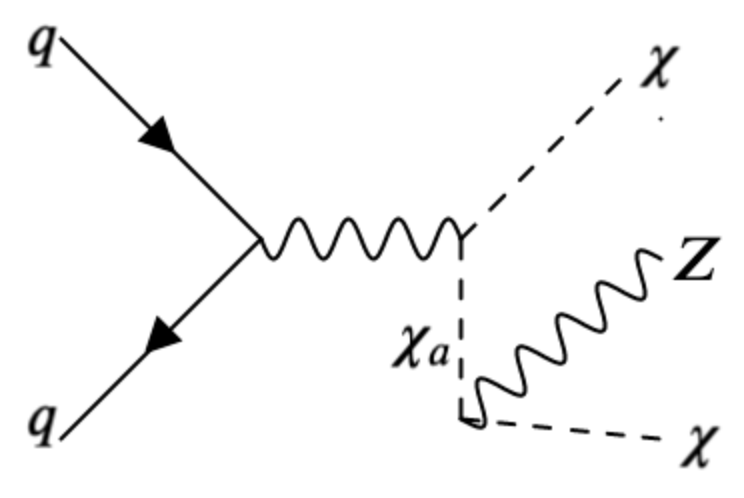}
\hspace{.034mm}
\includegraphics[scale = 0.34] {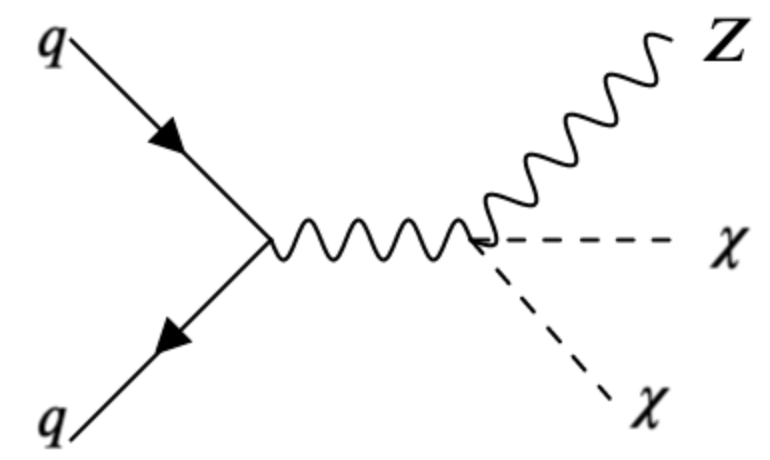}
\caption{ Feynman diagrams contributing to the mono Z process.} \label{monoZdiagrams}
\end{figure}

\begin{figure}[H]
\centering
\includegraphics[scale = 0.5] {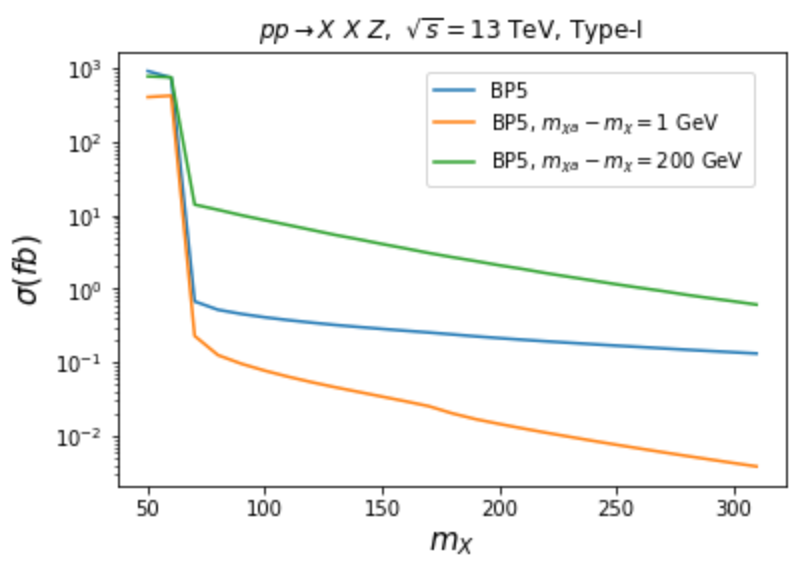}
\caption{ Cross sections for the mono-Z object final state for  BP5 of table \ref{monojet_XX_benchmarks}.   The parameter $m_{\chi_a}$ was varied according to different mass separations as shown in the legends. } \label{monoZXX}
\end{figure}

Another interesting signal that arises within this model is the mono Z final state. The relevant diagrams for this process are displayed in figure \ref{monoZdiagrams}. There are Higgs mediated diagrams which are determined by the $\lambda_{abc}$ coupling and are displayed in the first row of figure \ref{monoZdiagrams}.  
The interaction between the active heavy states and a $Z$ boson scales as $ g_{AHZ} \propto \sin{(\beta - \alpha)}$ while the heavy Higgs coupling to DM is proportional to $\cos{(\beta - \alpha)}$, therefore away from the alignment limit the pseudoscalar and heavy Higgs contribute to the cross section as shown in the third diagram of the first row. 
 This effect is however insignificant and the total cross section is dominated by the diagrams of the second row. This type of diagrams (second row) do not involve Higgs bosons and are completely determined by gauge interactions with $m_{\chi a}$ as the only relevant parameter.
 In figure \ref{monoZXX} we show the cross sections as a function of DM mass with other parameters fixed to the values corresponding to BP5 which has the highest deviation from alignment limit and maximum value of $\lambda_{abc}$.

\subsection{Mono-Higgs}

\begin{figure}[H]
\centering
\includegraphics[scale = 0.34] {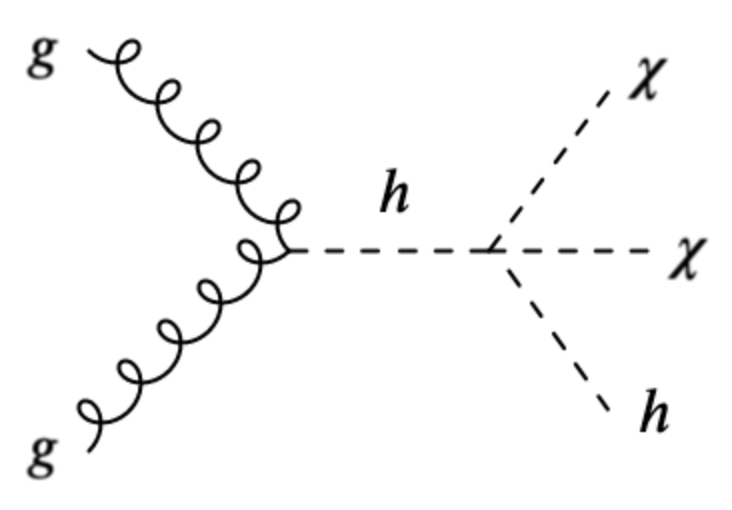}
\hspace{.644mm}
\includegraphics[scale = 0.34] {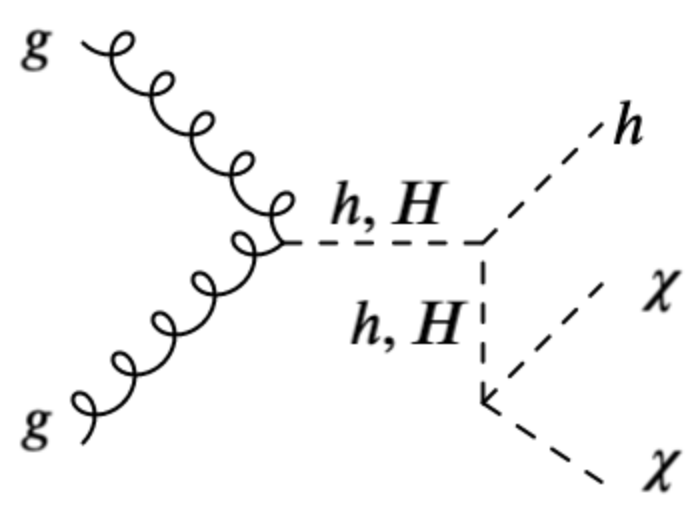}
\hspace{.044mm}
\includegraphics[scale = 0.34] {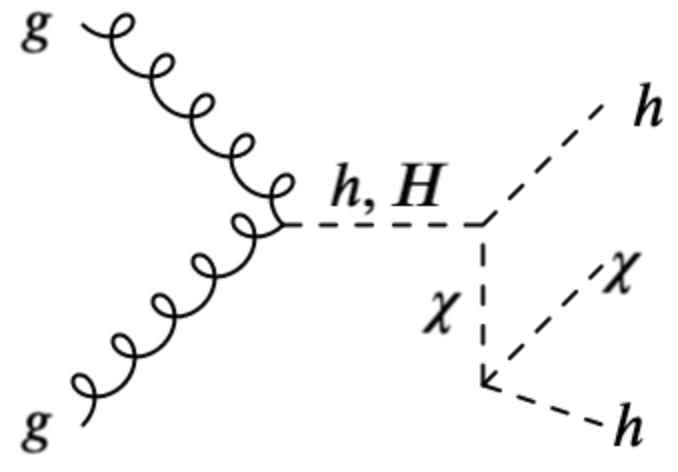}
\hspace{.044mm}
\includegraphics[scale = 0.34] {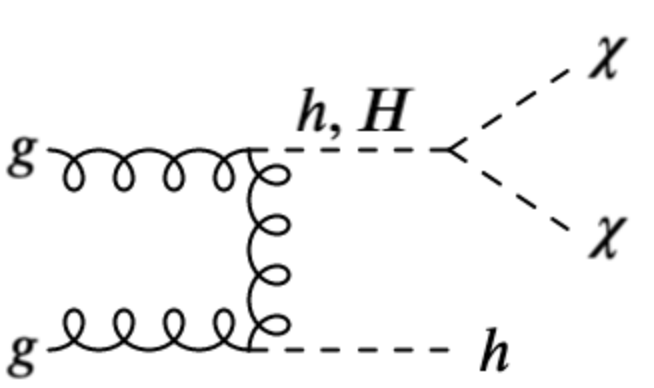}
\caption{ Feynman diagrams contributing to the mono-Higgs production via the process $p p \rightarrow \chi \chi h $.} \label{monohdiagrams}
\end{figure}

There is also the possibility of mono Higgs final states. Similar to the mono-jet production there are two possible final states, namely $h \chi \chi$ or $h \chi \chi_a$. Example diagrams for the former are displayed in figure \ref{monohdiagrams}. In this case the possibilities are more diverse having diagrams that scale with $g_{hhXX} = \lambda_{abc}$ as shown on the first diagram of the first row and also diagrams that scale with the trilinear Higgs couplings $g_{hhh}$, $g_{hhH}$ and $g_{hHH}$ as manifested in the second diagram of the first row. The Higgs trilinear couplings are given by

\begin{equation}
g_{hhh} = -\frac{1}{16 v c_{\beta}^2 s_{\beta}^2} \left[  -8 m_{12}^2 c_{\beta - \alpha}^2 c_{\beta + \alpha}  + 2 m_h^2  \left( c_{\beta -3\alpha} + 3 c_{\beta + \alpha} \right)s_{2\beta} \right],
\end{equation}
\begin{equation}
g_{hhH} = \frac{c_{\beta - \alpha}}{ v s_{\beta} c_{\beta}} \left[  m_{12}^2 \left(-1+3 \frac{c_{\alpha} s_{\alpha}}{c_{\beta} s_{\beta}} \right)  - \left( 2 m_h^2  + m_H^2 \right)s_{2\alpha} \right], \label{ghhH}
\end{equation}
\begin{equation}
g_{hHH} = \frac{8 s_{\beta - \alpha}}{ v s_{2\beta} } \left[ - m_{12}^2    +  s_{2\alpha} \left(  m_h^2  + 2 m_H^2 - 3\frac{m_{12}^2}{s_{2\beta}} \right) \right],
\end{equation}
where we used the shorthand notation $c_{\alpha} \equiv \cos{\alpha}$ etc. We thus see from these expressions that $m_{12}^2$ and $m_H$ determine the strength of the trilinear interactions for fixed $\alpha$ and $\beta$.  From the parameter scan we chose two benchmarks where this values are maximal and given by $g_{hhh}=-0.92 v$, $g_{hhH}=0.50 v$ and $g_{hHH}=3.40 v$ for BP7 and $g_{hhh}=-0.95 v$, $g_{hhH}=1.00 v$ and $g_{hHH}=-4.54 v$ for BP8. These benchmark points are presented in table \ref{monoh_XX_cs}.

  \begin{table}[h]
\begin{center}
\begin{tabular}{ | m{.9 cm} |m{1.cm}| m{1. cm}| m{.9 cm} | m{0.6 cm} | m{0.6 cm} |  m{0.9 cm} | m{0.6 cm} |m{0.905 cm} | m{0.6 cm} | m{1.1cm} | } 
\hline
 & $\tan{\beta}$ &  $c_{\beta - \alpha}$&$m_{12}$  & $m_H$ & $m_A$ & $m_{H \pm}$  & $m_{ \chi a}$ & $m_{ \chi \pm}$ & $\lambda_{\eta}$ & $\lambda_{abc}$  \\ 
\hline
BP7 & $4.44 $ & $ -0.36$& $20$ & $88$ & $224$ & $ 255$ & $847$  & $842$ &$8.23$ &$-0.11$ \\ 
\hline
BP8 & $6.51$ & $-0.20$& $102$ & $211$ & $386$ & $ 399$ &$733$ & $728$ &$7.15$ &$-0.16$ \\ 
\hline
\end{tabular}
\end{center} \caption{Benchmark points as drawn originally from the numerical scan which correspond to the curves of figure \ref{monohXX} .We took the square root of the absolute value of active sector mass parameter $m_{12}$ so that a negative value corresponds to $-m_{12}^2$ in the scalar potential. All mass parameters are in GeV units. } \label{monoh_XX_cs}
\end{table}

\begin{figure}[H]
\centering
\includegraphics[scale = 0.5] {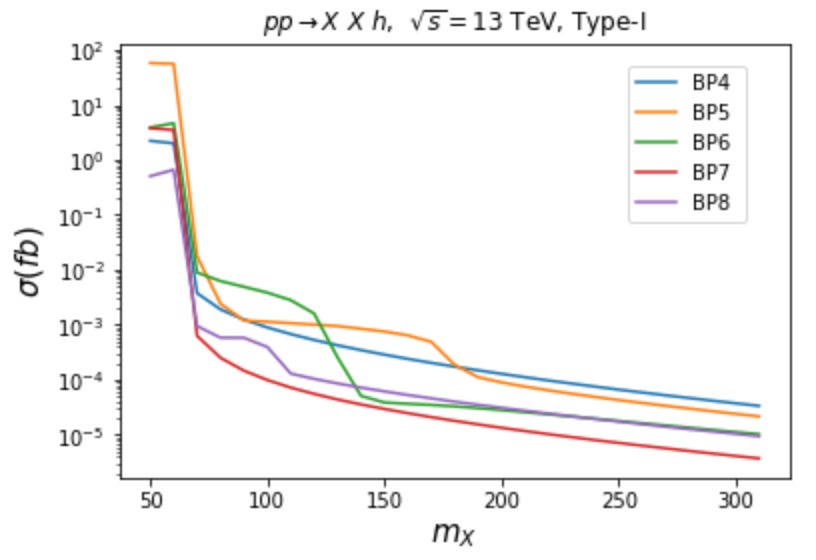}
\caption{Cross sections as a function of $m_{\chi}$ for mono-Higgs production via the $\chi \chi h$ final state. The benchmark points are presented in tables \ref{monojet_XX_benchmarks} and \ref{monoh_XX_cs}. } \label{monohXX}
\end{figure}

The results for the production cross sections are presented in figure \ref{monohXX} where for comparison we show the cross sections for the benchmarks BP4, BP5 and BP6. It can be seen that BP5, which corresponds to $\cos{(\beta - \alpha)} = 0.5$ produces the biggest cross section indicating non-trivial constructive interference effects away from the alignment limit. The cross section is dominant for large values of $|\lambda_{abc}|$ and is enhanced in the low mass region up to about the Higgs threshold when it starts to decrease.  
 
\begin{figure}[H]
\centering
\includegraphics[scale = 0.34] {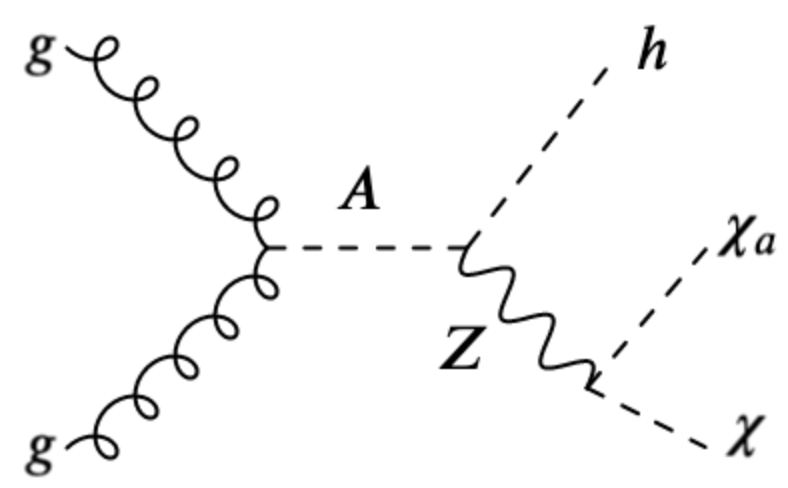}
\hspace{.644mm}
\includegraphics[scale = 0.34] {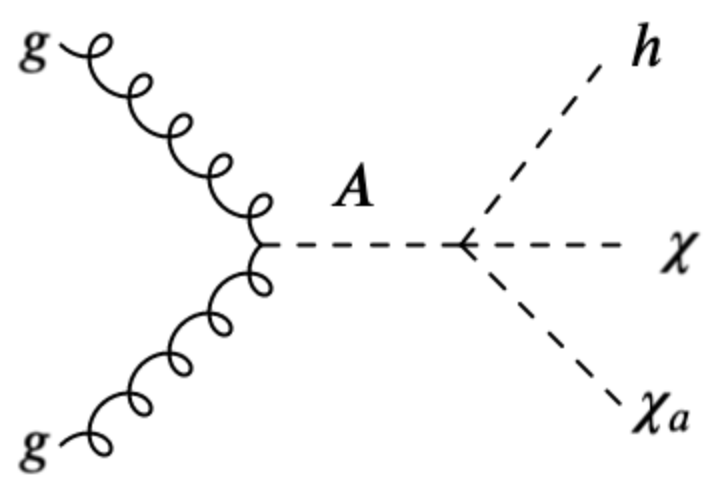}
\hspace{.044mm}
\includegraphics[scale = 0.34] {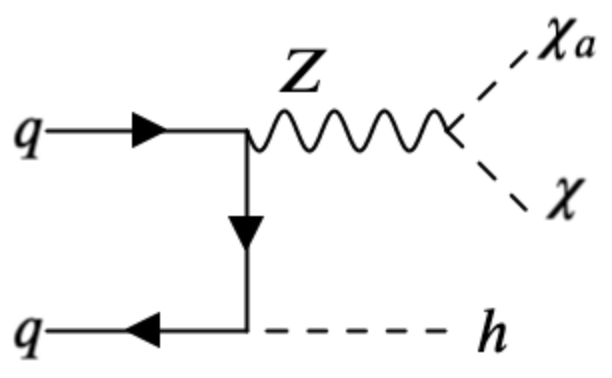}
\hspace{.044mm}
\includegraphics[scale = 0.34] {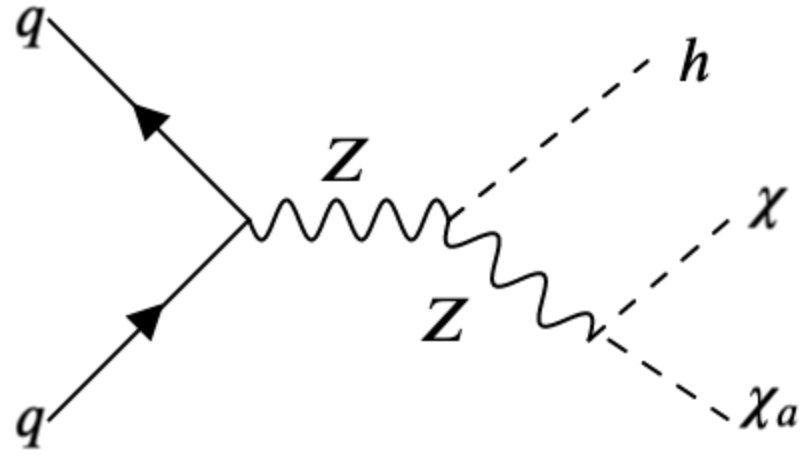}
\hspace{.044mm}
\includegraphics[scale = 0.34] {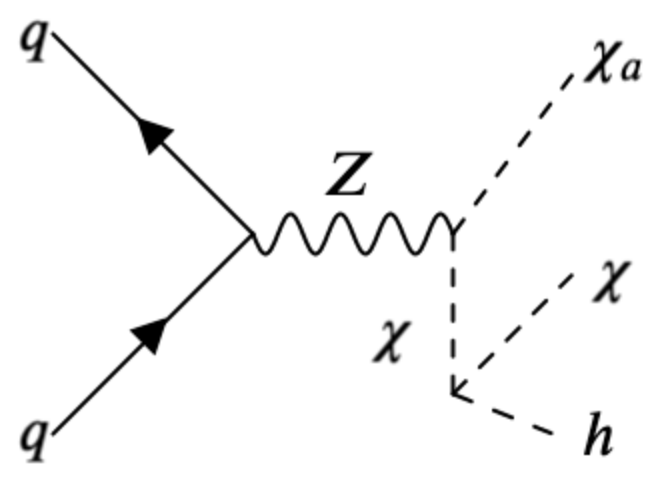}
\hspace{.044mm}
\includegraphics[scale = 0.34] {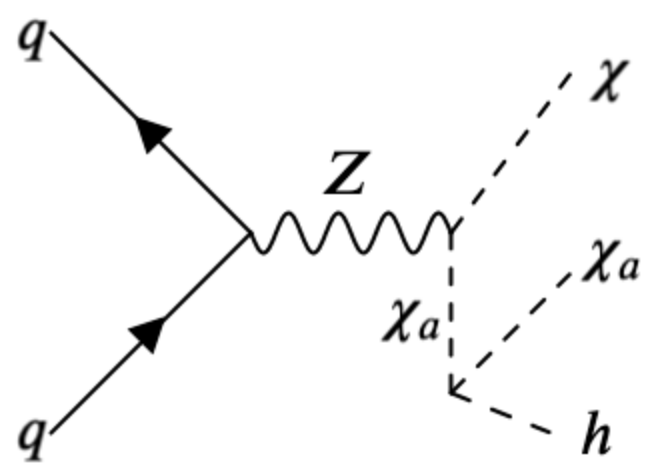}
\hspace{.044mm}
\includegraphics[scale = 0.34] {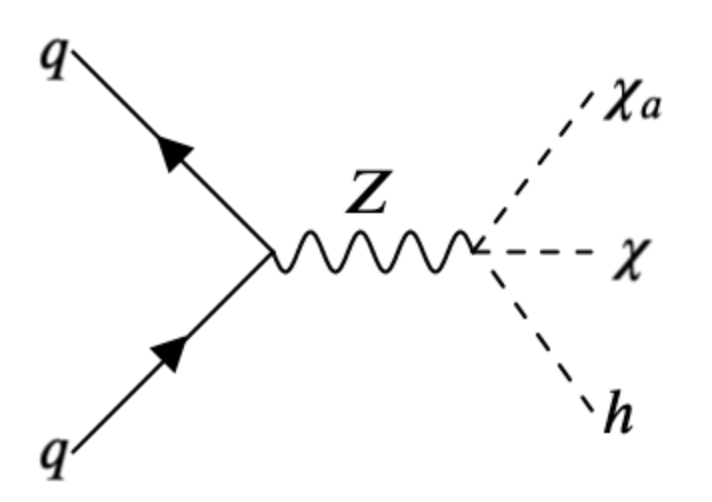}
\caption{Feynman diagrams contributing to the mono-Higgs production via the process $p p \rightarrow \chi \chi_a h $. } \label{monohXXadiagrams}
\end{figure}

\begin{figure}[H]
\centering
\includegraphics[scale = 0.5] {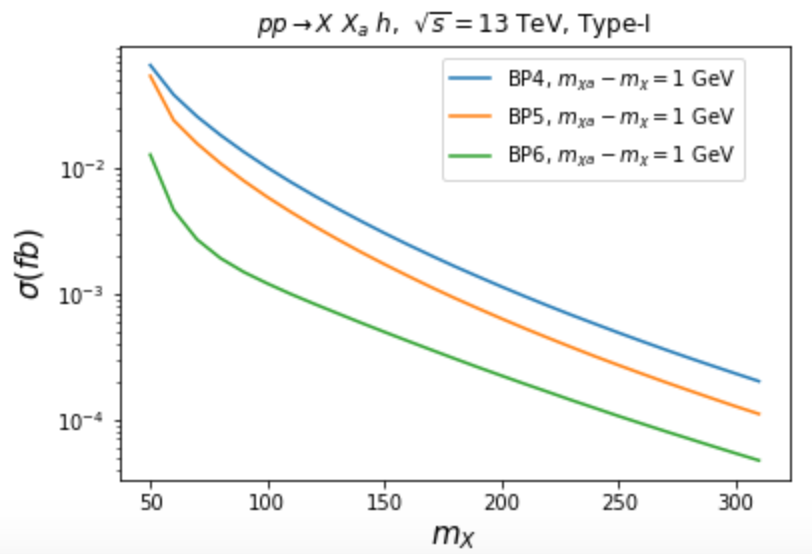}
\caption{Cross sections for $\chi \chi_a h $ final state as a function of DM mass for benchmark points BP4, BP5 and BP6. The mass parameter $m_{\chi_a}$ was varied such that the mass difference between $\chi_a$ and $\chi$ was kept fixed as shown in the legends.} \label{monohXXaplot}
\end{figure}

Finally, the Feynman diagrams for the process $p p \rightarrow h \chi \chi_a$ are shown in figure \ref{monohXXadiagrams}. The first two diagrams are mediated by the pseudoscalar and only contribute when $\cos{(\beta - \alpha)}\neq 0$. The biggest cross sections correspond to small mass separation between $m_{\chi_a}$ and $m_{\chi}$. 
We present plots for the cross sections in figure \ref{monohXXaplot}. We can see that for $\cos{(\beta - \alpha)}\neq 0$ BP5 and BP6 yield cross sections reduced relative to the alignment limit corresponding to BP4.


\subsection{Discussion}
From the cross sections presented above we can conclude that the strongest effect comes from the mono-jet final state $j \chi \chi$ with cross sections of about $\mathcal{O}(1)$ pb for a DM mass in the range $ 50 \leq m_{\chi} \leq 70$ GeV.

To test these predictions against the LHC data we used the CheckMATE 2  \cite{Dercks:2016npn,deFavereau:2013fsa,Cacciari:2011ma,Cacciari:2005hq,Cacciari:2008gp,Read:2002hq,Sjostrand:2014zea} software package. We implemented a grid search for different benchmark points in the plane of $\lambda_{abc}$ and $m_{\chi}$ in the range $ 1\leq \lambda_{abc} \leq 5$ and  $ 50 \leq m_{\chi} \leq 100$ GeV. For each point in the grid we generated partonic events using CalcHEP \cite{Belyaev:2012qa} with the same cuts and scales as specified at the beginning of this section. 

Parton showering and hadronization using Pythia 8  \cite{Sjostrand:2014zea}  was performed within CheckMATE 2. The program then performs a fast detector simulation using DELPHES 3 \cite{deFavereau:2013fsa}. We found all points in the grid to be allowed at $95 \%$ CL with the most sensitive analysis given in Ref. \cite{Aaboud:2017phn} which corresponds to an ATLAS search for DM using $36.1 \ fb^{-1}$ of data at $13 $ TeV center of mass energy. 
 
  This implies that, using mono-jet final states, current LHC data or at least the analyses currently implemented within CheckMATE 2 cannot rule out the I(1+2)HDM even for maximal values of $\lambda_{abc}$ allowed by theoretical and experimental constraints. We checked that the LDM and HDM regions are trivially allowed as they lead to very small cross sections either due to very small $\lambda_{abc}$ in LDM or very high mass in HDM. This effect can be better appreciated in figure \ref{LDM_HDM} (left) where we show the region in the $\lambda_{abc}$, $m_{\chi}$ plane. On the right of that figure we show the zoomed LDM region where the points colored pink are excluded by the projected bounds from LZ collaboration.   
 
  Our results agree with those of the IDM, see Ref. \cite{Belyaev:2018ext}. In that reference the authors obtained upper limits for current and projected luminosities by doing a shape analysis of the missing transverse momentum distribution.   Only by combining the final states $j \chi \chi + \ j \chi \chi_a$ for small mass separation $m_{\chi_a} = m_{\chi }+ 1$ GeV, they found that DM masses very close to the Higgs threshold $m_h/2$ could be excluded with $30 \ \text{fb}^{-1}$ of integrated luminosity. Further with $3000 \ \text{fb}^{-1}$ all the region with $m_{\chi} < m_h/2$ could be excluded for maximal Higgs-DM coupling. 

 The I(1+2)HDM stands as a simple extension of the IDM with more parameters that can weaken the limits coming from mono-jet searches at the LHC.

 \section{Dark Democracy Lifted } \label{NOT-DD}

\begin{figure}[H]
\centering
\includegraphics[scale = 0.5] {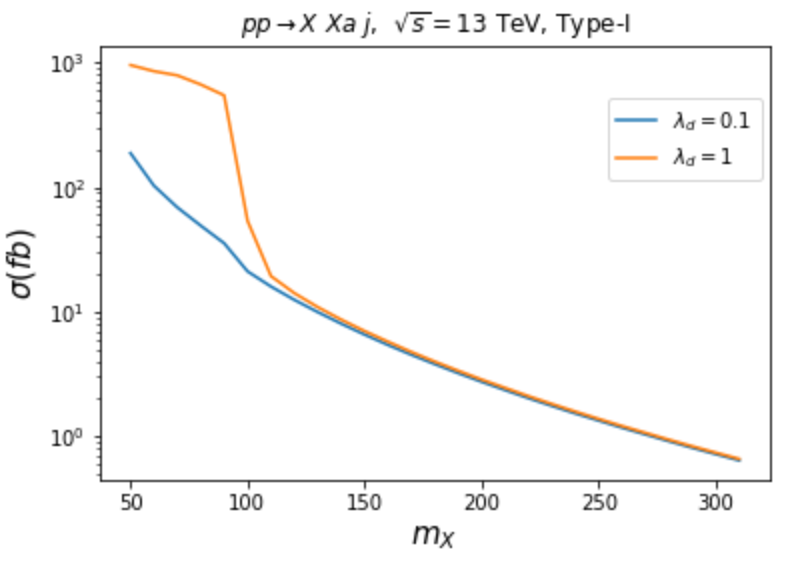}
\caption{Cross sections for the process $p p \rightarrow j \ \chi \chi_a $ as a function of DM mass for $\tan{\beta}=2$ and $m_A = 200$ GeV. The mass parameter $m_{\chi_a}$ was varied such that the mass difference was fixed to $m_{\chi_a} -m_{ \chi} = 1$ GeV.} \label{monohXXanotDD}
\end{figure}

In this section we explore the consequences of slightly relaxing the assumption of dark democracy of quartic couplings by taking $\lambda_{1313} \neq \lambda_{2323}$ and calling
\begin{equation}
\lambda_c \equiv  \lambda_{2323},
\end{equation}
\begin{equation}
\lambda_d \equiv \lambda_{1313} - \lambda_{c},
\end{equation}
with $\lambda_a$ and $\lambda_b$ as defined in \eqref{Eq:DarkDemocracy}. In this case, once again we can solve for the quartic couplings in favor of the scalar mass parameters and one obtains 
\begin{equation}
\lambda_a = \frac{2 (m_{\chi \pm}^2 - m_{\eta}^2)}{v^2},
\end{equation}
\begin{equation}
\lambda_b = \frac{m_{\chi_a}^2 - 2 m_{\chi{\pm}}^2 + m_{\chi}^2}{v^2},
\end{equation}
\begin{equation}
\lambda_c =  \frac{m^2_{\chi} - m^2_{\chi_a} }{v^2} - \lambda_d\cos^2{\beta},
\end{equation}
with $\lambda_a$ and $\lambda_b$ given by the same expressions as before. For this scenario the active set of parameters is modified to include the extra quartic coupling $\lambda_d$
\begin{equation}
S'_2 = \left\{  m_{\chi}, m_{\chi_a}, m_{\chi^{\pm}}, m_{\eta}^2, \lambda_{\eta}, \lambda_{d} \right\}, \label{set2}
\end{equation}
in this case the pseudoscalar would have a non vanishing coupling to pairs of inert states as
\begin{equation}
g_{A \chi \chi_a}  = \lambda_d v \cos{\beta} \sin{\beta}.
\end{equation}
As a byproduct of this non-vanishing coupling, the process $p p \rightarrow j \ \chi \ \chi_a$ will also be mediated by a virtual pseudoscalar $A$ that decays into $\chi  \ \chi_a$ modifying the cross sections that we calculated for the Dark democracy case in the preceding section. The extra diagrams with a pseudoscalar mediator thus depend on $m_A$, $\tan{\beta}$ and the new quartic $\lambda_d$.  We present the cross sections in figure \ref{monohXXanotDD} for $\lambda_d=0.1 $ and $\lambda_d=1 $ with $\tan{\beta}=2$ and $m_A =200$ GeV fixed. We notice that there is a  significant enhancement relative to the democratic case of about one order of magnitude if $\lambda_d = 1$.  
As noted in Section 7.1, the current experimental limits on the process are approximately one or two orders of magnitude above the expected value in this model, so a large value of $\lambda_d$ will bring them closer.   Certainly, the HL-LHC will cover a substantial part of the region.

\section{Heavy Higgs decays} \label{HeavyHiggs}

Constraints on the parameter space of 2HDMs with softly broken $Z_2$ symmetry due to experimental heavy Higgs searches at the LHC have been studied before, see e.g. Ref. \cite{Chowdhury:2017aav}.  Here, we present  predictions for production cross sections times branching fractions using a benchmark value of the parameters and we will compare with experimental searches to indicate if the model can be probed or not.

We focus on the decay mode $H \rightarrow hh$ for concreteness.  The production cross section depends on the heavy Higgs couplings to quarks and on the trilinear coupling $g_{hhH}$ given in eq. \eqref{ghhH}. From the parameter scan we chose a benchmark value where the decay width has its maximum value. This benchmark is presented in table \ref{benchmarks_H} below. 

  \begin{table}[h]
\begin{center}
\begin{tabular}{ | m{.9 cm} |m{1.cm}| m{1. cm}| m{.9 cm} | m{0.6 cm} | m{0.6 cm} |  m{0.9 cm} | m{0.6 cm} |m{0.605 cm} | m{0.6 cm} | m{0.9cm} |m{0.9cm} | } 
\hline
 & $\tan{\beta}$ &  $c_{\beta - \alpha}$&$m_{12}$  & $m_H$ & $m_A$ & $m_{H \pm}$ & $m_{\eta}$ & $m_{\chi}$ & $m_{ \chi a}$ & $m_{ \chi \pm}$ & $\lambda_{\eta}$   \\ 
\hline
BP9 & $3.35 $ & $ 0.51$& $512$ & $693$ & $603$ & $ 488$& $366$ & $372$ & $379$  & $375$ &$5.62$  \\ 
\hline
\end{tabular}
\end{center} \caption{Benchmark point as drawn originally from the numerical scan which correspond to the curve of figure \ref{pp_H_hh} . All mass parameters are in GeV. } \label{benchmarks_H}
\end{table}

\begin{figure}[H]
\centering
\includegraphics[scale = 0.5] {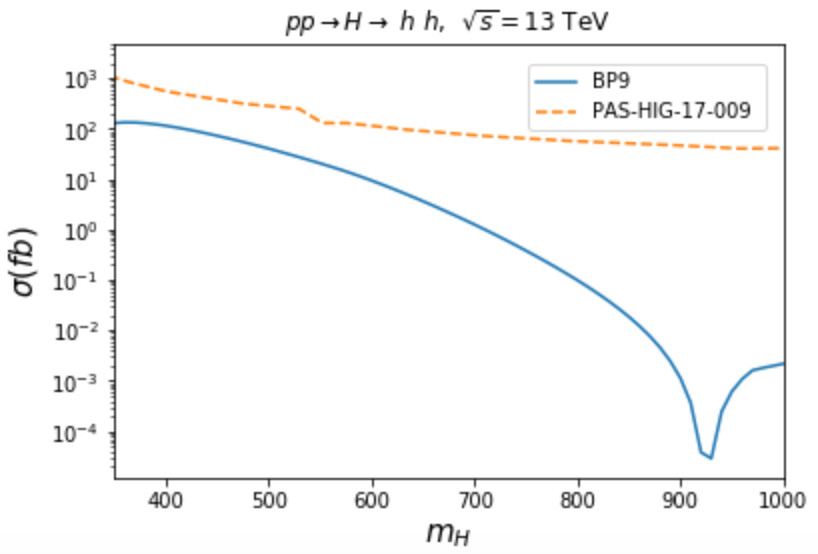}
\caption{Production cross section times branching ratios for the process $p p \rightarrow H \rightarrow hh (bbbb) $ as a function of $m_H$ for the benchmark BP9 given in table 4. We also show the 95$\%$  CL observed upper limit from Ref. \cite{CMS:2017rli}.   } \label{pp_H_hh}
\end{figure}

The magnitude of the observed experimental upper bounds depends on the decay mode of the final state Higgs bosons of the respective search analysis. In figure 11 of Ref. \cite{CMS:2017xxp} upper bounds for the production cross section times branching fraction are presented corresponding to different final states of each SM Higgs boson. One can see that the strongest constraints are given by searches focused on the $bbbb$ final state. As no signal excess was observed in the experiment we plot the 95 $\%$ confidence level upper limit (observed)
 on the cross section in figure \ref{pp_H_hh} together with the cross section for benchmark BP9. We notice that heavy Higgs searches on this decay mode can not test the cross section. 
  
 One clear distinction of the I(1+2)HDM is the possibility of heavy Higgs decays into inert scalars, a possibility that exists in neither the usual 2HDM nor the usual IDM.   This is an invisible decay that would have the effect of suppressing the branching ratio to other states.    Note that many effects of heavy Higgs bosons are included in the previous section, including the effects of invisible decays in mono-jet, mono-Z and mono-Higgs searches.    We now focus on the suppression of the branching ratios of heavy scalars.

Since the invisible decay of the $H$ depends on the $H\chi \chi$ vertex, which is $\lambda_{abc}v\cos(\beta-\alpha)$, and since the $HZZ$ vertex also depends on $\cos(\beta-\alpha)$, the ratio of the two decays will be independent of the mixing angles.    For typical values of $\lambda_{abc}$ given in table \ref{monojet_XX_benchmarks}, the ratio of the decays to $ZZ$ vs. into $\chi\chi$ varies from roughly $0.5$ to $1$, ignoring phase space.    This will suppress the effective branching ratio to vector bosons by roughly a factor of two at most.     Given that we appear to be near the alignment limit, at which both decays will vanish, the decay into $\tau^+\tau^-$ may become critical (the decays into $\bar{b}b$ and $\bar{c}c$ are much more difficult to see due to backgrounds).    The ratio of $H\rightarrow\chi\chi$ to $H\rightarrow\tau\tau$ is (neglecting phase space)
\begin{equation}
\frac{\Gamma(H\rightarrow\chi\chi)}{\Gamma(H\rightarrow\tau\tau)} = \frac{\lambda_{abc}^2 v^4}{4m_H^2m_\tau^2}\frac{\cos^2(\beta-\alpha)\sin^2\beta}{\sin^2\alpha}
\end{equation}
For typical values of $\lambda_{abc}$ in Table \ref{monojet_XX_benchmarks}, and for an $H$ mass of $200$ GeV, this is approximately $3\cos^2(\beta-\alpha)\sin^2\beta/\sin^2\alpha$.  This could be reduced by phase space, of course.    In the alignment limit, this vanishes, but if one is at the larger values of $\cos(\beta-\alpha)$ allowed in type I models, it can be substantial.  Note that one could make the ratio dominate in the limit of $\alpha$ very small, but that is a region of parameter-space in which the $H\rightarrow\tau\tau$ decay is unmeasurable.       Nonetheless, this invisible decay could suppress a promising signature of the $H$.

For the pseudoscalar, $A$, the $A\rightarrow\tau\tau$ signature is more promising, even for larger masses, since the decay into $W$'s and $Z$'s is absent.   However, as noted in the last section, the decay $A\rightarrow\chi\chi_a$ vanishes in the "dark democracy" limit, thus invisible decays would be absent.    If one does not adopt dark democracy, then there will be a signal.   Defining $\lambda_d$ as in the last section, the ratio of invisible to tau decays is
\begin{equation}
\frac{\Gamma(A\rightarrow \chi\chi_a)}{\Gamma(A\rightarrow\tau\tau)}= \frac{\lambda_d^2\sin^4\beta}{2(m_\tau/v)^2}
\end{equation}
times the usual phase space factors.    Unless one is very close to the dark democracy limit, this expression will be quite large, and if kinematically allowed, the invisible decay of the $A$ will dominate, swamping the visible decay.

If the $\chi$ is the dark matter particle, then $\chi_a$ will be heavier, and the decay is only kinematically allowed for $m_A$ greater than about $145$ GeV.   Above a mass of $210$ GeV, the decay into $h + Z$ can be important (away from the alignment limit),  and thus the range of masses in which the invisible decay is dominant may be small.   Nonetheless, a study of invisible decays in this case could be interesting.

\section{Conclusions} \label{conclusions}

In this paper we have investigated the constraints on the parameter space of the I(1+2)HDM. A random scan of the free parameters was performed taking into account positivity of the potential and unitarity of the quartic couplings as theoretical constraints. Experimental constraints such as B physics, EWPO, LEP bounds on gauge bosons decays, LHC data on the SM Higgs boson, heavy Higgs searches, upper limit of the DM relic density and current direct detection constraints on DM-nucleon scattering have been imposed.

Two regions that satisfy also the lower limit on the relic density and thus are non under-abundant have been identified. There is a low mass region we called LDM with $m_{\chi} \in [57,73] $ GeV and a high mass region HDM with $m_{\chi} \in [500,1000] $ GeV. This result is consistent with previous studies of this model \cite{Grzadkowski:2010au}. For a given value of $m_{\chi}$, the Higgs coupling to DM is controlled by the inert sector mass squared parameter as shown in equation \ref{DMcoupling}.

 The projected sensitivity of the LZ experiment has the potential to exclude a significant amount of parameter space leaving a maximum separation of $|m_{\eta} - m_{\chi}| \leq 0.1$ GeV in LDM and $|m_{\eta} - m_{\chi}| \leq 0.25$ GeV in HDM.
  For the LDM there is a tiny window that would survive with $70 \leq m_{\chi} \leq 73$ GeV and quartic coupling values of about $\lambda_{abc} \sim \mathcal{O}(10^{-4})$.   Of course, in any parameter scan, one can miss ``funnel" regions where different processes cancel; we have not considered the possibility of this fine-tuning.

The maximum deviation from the alignment limit in LDM and HDM regions was found to be in agreement with the overall shape of the parameter space of mixing angles of type-I and type-II models \cite{Haller:2018nnx}. This demonstrates that relic density constraints are only dependent on the quartic $\lambda_{345}$ as is evident from figure \ref{alpha_beta}.

 Predictions of the model for mono-jet, mono Z and mono Higgs final states have been studied. For well motivated benchmarks it has been shown that the most competitive signal is given by $ p p \rightarrow j \chi \chi$ with cross sections of about $\mathcal{O}(1)$ pb for DM mass in the range $50 < m_{\chi} < 70$ GeV and a Higgs DM interaction of about $\lambda_{abc} \approx 0.2$. The model has been tested using CheckMATE 2 and we found that it is allowed at 95 $\%$ CL by the LHC analyses implemented in this package. 
 
We then considered the ``dark democracy" assumption common in IDM studies.   This assumption is often made to simplify the parameter-space.   However in this model (unlike either the IDM or 2HDM models), the process $A\rightarrow\chi\chi_a$ can occur once the dark democracy assumption is lifted.   This decay would be a convincing signature of the model, and can lead to a significant enhancement of the mono-jet cross section.
 
We have shown that searches for DM at the LHC in final states with a jet offer a difficult way to test this model however future direct detection experiments will be able to challenge this scenario as a model that can account for all the DM in the universe.     As is the case in the IDM, the current LHC bounds on mono-particle processes  are not sufficient to test the model, but the HL-LHC will, after $3000\ {\rm fb}^{-1}$, be able to probe a substantial part of the parameter space.

\begin{acknowledgments}
After the completion of this work we became aware that the I(1+2)HDM was being studied in the context of the minimal extended seesaw framework in Ref. \cite{Das:2019ntw}. We thank Najimuddin Khan for pointing us to this reference. MM thanks  Alexander Belyaev and Igor Ivanov for helpful correspondence. The authors would also like to thank Eloy Romero Alcalde for helping with the installation of CheckMATE 2.  This work was supported by the National Science Foundation under  Grant PHY-1819575.

\end{acknowledgments}

\begin{appendices}
\section{2HDM parameters}\label{2HDM}

The minimization conditions on the potential $V_{12}$ allows us to solve for the quadratic mass terms as
\begin{equation}
m_{11}^2 = v^2 (\lambda_1 \cos^2{\beta} + \lambda_{345} \sin^2{\beta}) - 
 m_{12}^2 \tan{\beta},
\end{equation}
\begin{equation}
m_{22}^2 = -m_{12}^2 \cot{\beta} +  v^2 (\lambda_{345} \cos^2{\beta} + \lambda_2 \sin^2{\beta}),
\end{equation}
where we define
\begin{equation}
\lambda_{345} \equiv \lambda_3 +\lambda_4 +\lambda_5.
\end{equation}

The mass squared matrix for the CP-odd and charged Higgs bosons are given by
\begin{equation}
M_{\text{Odd}}^2 = 
  \begin{pmatrix}
1/2 (m_{12}^2 - v^2 \lambda_5 \sin{2 \beta}) \tan{\beta} & -m_{12}^2/2 + v^2 \lambda_5 \cos{\beta} \sin{\beta}  \\
-m_{12}^2/2 + v^2 \lambda_5 \cos{\beta} \sin{\beta}   & 1/2 \cot{\beta} (m_{12}^2 - v^2 \lambda_5 \sin{2 \beta})
\end{pmatrix} ,
\end{equation}
\begin{equation}
M_{\text{Charged}}^2 = 
  \begin{pmatrix}
-v^2 (\lambda_4 + \lambda_5) \sin^2{\beta} + m_{12}^2 \tan{\beta} & -m_{12}^2 + v^2 (\lambda_4 + \lambda_5) \cos{\beta} \sin{\beta} \\
-m_{12}^2 + v^2 (\lambda_4 + \lambda_5) \cos{\beta} \sin{\beta}  & -v^2 (\lambda_4 + \lambda_5) \cos^2{\beta} + m_{12}^2 \cot{\beta}
\end{pmatrix} ,
\end{equation}
respectively. Diagonalization of this matrices, by eq. \ref{diag1},  yield a zero eigenvalue corresponding to the Goldstone bosons that gets eaten to become the $Z$ and $W$ bosons longitudinal polarizations. We choose to solve for the quartic couplings in favor of the pseudoscalar and charged scalar masses squared $m_{A}^2$, $m_{H^{\pm}}^2$ as 
\begin{equation}
\lambda_5  = \frac{-m_{A}^2 + m_{12}^2 \csc{2 \beta}}{v^2},
 \end{equation}
\begin{equation}
\lambda_4 = \frac{ m_{A}^2 - 2 m_{H^{\pm}}^2 + m_{12}^2 \csc{2 \beta}}{v^2}.
\end{equation}

The CP-even mass squared matrix is given by 
\begin{equation}
M_{\text{Even}}^2 = 
  \begin{pmatrix}
v^2 \lambda_1 \cos^2{\beta} +  m_{12}^2 \tan{\beta}/2 & -m_{12}^2/2 + v^2 \lambda_{345} \cos{\beta} \sin{\beta}  \\
-m_{12}^2/2 + v^2 \lambda_{345} \cos{\beta} \sin{\beta}  &  m_{12}^2 \cot{\beta} /2+ v^2 \lambda_2 \sin^2{\beta}
\end{pmatrix} ,
\end{equation}
where the angle that diagonalizes this matrix is given by the formula
\begin{equation}
\tan{2 \alpha}   =  \frac{2 M_{\text{Even},12}^2}{M_{\text{Even,11}}^2-M_{\text{Even,22}}^2}.
\end{equation}

We use the condition 
\begin{equation}
R(-\alpha)M^2_{\text{Even}}R(\alpha)  = \begin{pmatrix}
m_H^2& 0 \\
0 & m_h^2
\end{pmatrix},
\end{equation}
 where $R(\alpha)$ is the rotation matrix of equation \eqref{scalars} to solve to solve for $\lambda_1$ and $\lambda_2$ and the off diagonal element to solve for $\lambda_{345}$. They are given by 
\begin{equation}
\lambda_1 = \frac{m_H^2 \cos^2{\alpha}+ m_h^2 \sin^2{\alpha}-1/2 \ m_{12}^2 \tan{\beta}}{v^2 \cos^2{\beta}},
\end{equation}
\begin{equation}
\lambda_2 = \frac{m_H^2 \sin^2{\alpha}+ m_h^2 \cos^2{\alpha}-1/2 \ m_{12}^2 \cot{\beta}}{v^2 \sin^2{\beta}},
\end{equation}
\begin{equation}
\lambda_{345} = \frac{m_{12}^2 - (m_h^2-m_H^2)\sin{2\alpha}}{2v^2 \cos{\beta}\sin{\beta}}.
\end{equation}

\section{EWPO Formulas }  \label{EWPOformulas}
\begin{align}
\Delta S_{\text{A}}& = \frac{1}{4\pi }\Bigg\{  s_{\beta-\alpha}^2F'(m_Z^2;m_H,m_A)-F'(m_Z^2;m_{H^\pm},m_{H^\pm}) \notag\\
& +c_{\beta-\alpha}^2\Big[F'(m_Z^2;m_h,m_A)+F'(m_Z^2;m_H,m_Z)-F'(m_Z^2;m_h,m_Z)\Big]\notag\\
& +4m_Z^2c_{\beta-\alpha}^2\Big[G'(m_Z^2;m_H,m_Z)-G'(m_Z^2;m_h,m_Z)\Big]\Bigg\}, \\
\Delta T_{\text{A}}& = \frac{1}{16\pi^2 \alpha_{\text{em}} v^2}
\Bigg\{ 
F(0;m_{H^\pm},m_A)
+s_{\beta-\alpha}^2[F(0;m_{H^\pm},m_H)-F(0;m_A,m_H)]\notag\\
& +c_{\beta-\alpha}^2\Big[F(0;m_{H^\pm},m_h)+F(0;m_H,m_W)+F(0;m_h,m_Z)\notag\\
&\quad\quad\quad -F(0;m_h,m_W)-F(0;m_A,m_h)-F(0;m_H,m_Z)\notag\\
&\quad\quad\quad + 4G(0;m_H,m_W)+4G(0;m_h,m_Z)-4G(0;m_h,m_W)-4G(0;m_H,m_Z)\Big]  \Bigg\}, \\
\Delta U_{\text{A}}& = \frac{1}{4\pi}
\Bigg\{ 
F'(m_W^2;m_{H^\pm},m_A)-F'(m_Z^2;m_{H^\pm},m_{H^\pm})\notag\\
&\quad\quad+s_{\beta-\alpha}^2[F'(m_W^2;m_{H^\pm},m_H)-F'(m_Z^2;m_A,m_H)]\notag\\
&\quad\quad +c_{\beta-\alpha}^2\Big[F'(m_W^2;m_{H^\pm},m_h)+F'(m_W^2;m_W,m_H)-F'(m_W^2;m_W,m_h)\Big]  \notag\\
&\quad\quad -c_{\beta-\alpha}^2\Big[F'(m_Z^2;m_A,m_h)+F'(m_Z^2;m_Z,m_H)-F'(m_Z^2;m_Z,m_h)\Big]  \notag\\
&\quad\quad +4m_W^2c_{\beta-\alpha}^2\Big[G'(m_W^2;m_H,m_W)-G'(m_W^2;m_h,m_W)\Big]\notag\\
&\quad\quad -4m_Z^2c_{\beta-\alpha}^2\Big[G'(m_Z^2;m_H,m_Z)-G'(m_Z^2;m_h,m_Z)\Big] \Bigg\}, \\
\Delta S_{\text{I}}&=\frac{1}{4\pi}
\Big[F'(m_Z^2;m_{\eta_H^{}},m_{\eta_A^{}}) - F'(m_Z^2;m_{\eta^\pm},m_{\eta^\pm})\Big],\\
\Delta T_\text{I}
&=
\frac{1}{16\pi^2 \alpha_{{\rm em}} v^2} 
\Big[F(0;m_{\eta^\pm},m_{\eta_A^{}}) 
+ F(0;m_{\eta^\pm},m_{\eta_H^{}})-F(0;m_{\eta_A^{}},m_{\eta_H^{}})  \Big],\\
\Delta U_\text{I}&=\frac{1}{4\pi}
\Big[F'(m_W^2;m_{\eta^\pm},m_{\eta_H^{}}) + F'(m_W^2;m_{\eta^\pm},m_{\eta_A^{}})\notag\\
&-F'(m_Z^2;m_{\eta^\pm},m_{\eta^\pm}) -F'(m_Z^2;m_{\eta_H},m_{\eta_A})   \Big],
\end{align}
where $F'(m_V^2;m_1,m_2)=[F(m_V^2;m_1,m_2)-F(0;m_1,m_2)]/m_V^2$ and $G'(m_V^2;m_1,m_2)=[G(m_V^2;m_1,m_2)-G(0;m_1,m_2)]/m_V^2$. 
The loop functions are given by 
\begin{align}
F(p^2;m_1,m_2)&=\int_0^1 dx \Big[(2x-1)(m_1^2-m_2^2)+(2x-1)^2p^2\Big]\ln \Delta_B, \\
F(0;m_1,m_2)&=\frac{1}{2}(m_1^2+m_2^2)+\frac{2m_1^2m_2^2}{m_1^2-m_2^2}\ln\frac{m_2}{m_1}, \\
G(p^2;m_1,m_2)&= \int_0^1 dx  \ln \Delta_B , \\
G(0;m_1,m_2)&=\ln(m_1m_2)-\frac{m_1^2+m_2^2}{m_1^2-m_2^2}\ln\frac{m_2}{m_1}-1, \\
\Delta_B&=  xm_1^2+(1-x)m_2^2-x(1-x)p^2. 
\end{align}

\end{appendices}

\providecommand{\href}[2]{#2}\begingroup\raggedright 

\endgroup

\end{document}